\documentclass{rQUF2e}
\usepackage{graphicx}
\usepackage{subfig}
\usepackage{mathtools}

\theoremstyle{plain}

\theoremstyle{definition}

\theoremstyle{remark}

\begin{document}


\title{Model risk on credit risk}

\author{J. MOLINS and E. VIVES${\dag}$\\
\affil{
$\dag$ \textit{Dept. d'Estructura i Constituents de la Mat\`eria, Universitat de Barcelona, Diagonal 647, Facultat de F\'isica, 08028 Barcelona, Catalonia (Spain)}
} \received{Draft released July 2015} }

\maketitle

\begin{abstract}
This paper develops the Jungle model in a credit portfolio framework. The Jungle model is able to model credit contagion, produce doubly-peaked probability distributions for the total default loss and endogenously generate quasi phase transitions, potentially leading to systemic credit events which happen unexpectedly and without an underlying single cause. We show the Jungle model provides the optimal probability distribution for credit losses, under some reasonable empirical constraints. The Dandelion model, a particular case of the Jungle model, is presented, motivated and exactly solved. The Dandelion model provides an explicit example of doubly-peaked probability distribution for the credit losses. The Diamond model, another instance of the Jungle model, experiences the so called quasi phase transitions; in particular, both the U.S. subprime and the European sovereign crises are shown to be potential examples of quasi phase transitions. We argue the three known sources of default clustering (contagion, macroeconomic risk factors and frailty) can be understood under the unifying framework of contagion. We suggest how the Jungle model is able to explain a series of empirical stylized facts in credit portfolios, hard to reconcile by some standard credit portfolio models. We show the Jungle model can handle inhomogeneous portfolios with state-dependent recovery rates. We look at model risk in a credit risk framework under the Jungle model, especially in relation to systemic risks posed by doubly-peaked distributions and quasi phase transitions.
\end{abstract}



\clearpage

{\abstractfont\centerline{\bfseries Index}\medskip
\hbox to \textwidth{\hsize\textwidth\vbox{\hsize19pc
\hspace*{-12pt} {1.}    Introduction\\
\hspace*{7pt} {1.1.}  Related literature\\
{2.}    The data\\
{3.}    Credit portfolio modelling\\
{4.}    The Jungle model and credit risk\\
{5.}  The Jungle model, hands on\\
\hspace*{7pt} {5.1.}  The binomial model\\
\hspace*{7pt} {5.2.}  Small contagion\\
\hspace*{7pt} {5.3.}  The Dandelion model\\
\hspace*{7pt} {5.4.}  The Diamond model\\
\hspace*{7pt} {5.5.}  The Jungle model and the real world\\
{6.}    The Jungle model and model risk\\
{7.}    Modelling inhomogeneous portfolios\\
\hspace*{8pt} and recovery rates\\
\hspace*{7pt} {7.1.} Modelling inhomogeneous portfolios,\\
\hspace*{8pt} no modelling for recovery rates\\
\hspace*{7pt} {7.2.} Homogeneous portfolios with\\
\hspace*{8pt} state-dependent recovery rates\\
\hspace*{7pt} {7.3.} Inhomogeneous portfolios\\
\hspace*{8pt} with state-dependent recovery rates\\
{8.}    Contagion, macroeconomic risk\\
\hspace*{8pt} factors and frailty\\
\hspace*{7pt} {8.1.} Macroeconomic risk factors\\
\hspace*{8pt}  as contagion\\
\hspace*{7pt} {8.2.} Frailty as contagion\\
{9.}    Policy implications of contagion\\
\hspace*{7pt} {9.1.}  The U.S. subprime and the European\\
\hspace*{8pt} sovereign crises as quasi-phase transitions\\
\hspace*{7pt} {9.2.}  Understanding the historical\\
\hspace*{8pt} probability distributions of credit losses\\
\hspace*{7pt} {9.3.}  How should the Jungle model \\
\hspace*{8pt} be used in practice?\\
\hspace*{7pt} {9.4.}  It's the correlations, stupid!\\
\hspace*{7pt} {9.5.}  "Too Big To Fail" banks\\
{10.}   Conclusions \\
}}}

\clearpage

\section{Introduction}

Clustering of corporate defaults is relevant for both macroprudential regulators and banks' senior management. With a robust modelling of credit losses, macroprudential regulators may analyse and manage the risk of systemic events in the economy, and banks' senior management may compute the capital needs out of their core credit portfolios.

Historical corporate default rate data, as described in \citep{moodys} and \citep{giesecke11}, signal the sensitivity of credit defaults to systemic events in the economy, from the Great Depression and the 2007-2009 Great Recession, to the savings and loans crisis and the burst of the dotcom bubble, as it can be seen from Figure 1.

\begin{figure}[!ht]
	\centering
	\includegraphics[width=135mm,height=85mm]{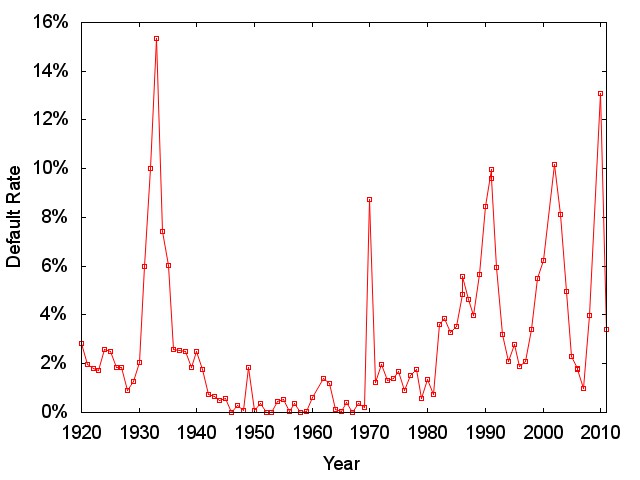}
	\caption{Historical default rates for Global Speculative-Grade bonds, from \citep{moodys} \label{overflow}}
\end{figure}

Standard credit portfolio models were not able to model the tail risks in credit portfolios when the U.S. subprime and the Spanish Real Estate bubbles bursted. Some of these models introduce default correlations through the dependency of the probabilities of default on macroeconomic factors describing the state of the economy. As a consequence, when the state of the economy is "good", the probabilities of default tend to go down. Conversely, when the state of the economy is "bad", the probabilities of default tend to go up. Averaging over the business cycle induces default clustering.

However, these predicted default correlations tend to be low in comparison to empirical evidence, and the corresponding probability distribution of the losses shows "thin tails". It is widely accepted that in addition to the dependence on macroeconomic risk factors, a reasonable credit risk model should include contagion effects, too. 

Contagion effects should often give rise to doubly-peaked probability distributions for the credit losses, with the first peak being close to the peak of an equivalent binomial distribution (when contagion effects are weak, and the defaults can be considered as roughly independent, which is usually the case when the state of the economy is "good") and a second peak, at higher losses, corresponding to avalanches / domino effects of credit defaults due to contagion.

This paper has the purpose to show a particular class of credit risk model, the Jungle model \footnote{The name "Jungle" provides intuition for the complex network of dependencies among the constituents of a credit portfolio. Moreover, since the lion is the King of the Jungle, we will see the Dandelion model (from the French "dent-de-lion", or lion's tooth) is the King of the Jungle of contagion models, since the Dandelion may describe the contagion arising from banks, which are the main source of systemic risks.}, is able to:

\begin{enumerate}
	\item Model contagion among borrowers
	\item Endogenously generate doubly-peaked probability distributions for the credit losses. As opposed to the case of single-peaked probability distributions, for which higher credit losses are always less likely than lower losses (at the large loss regime), doubly-peaked probability distributions show the distressing phenomenon that very large losses may be more likely to happen than moderately large losses
	\item Show how credit systemic events may occur suddenly and unexpectedly. A credit portfolio may inadvertently cross a "quasi phase transition point", and its collective behaviour change all of a sudden, potentially creating systemic events. We want to emphasize that intuition usually tells us a systemic crisis requires a strong single cause originating it; however, this is not necessarily true. We will show a systemic crisis can be created without a strong underlying, single cause, and we will learn how to recognize those "quasi phase transition points"
\end{enumerate}

\begin{figure}[ht!]
	\centering
	\includegraphics[width=95mm]{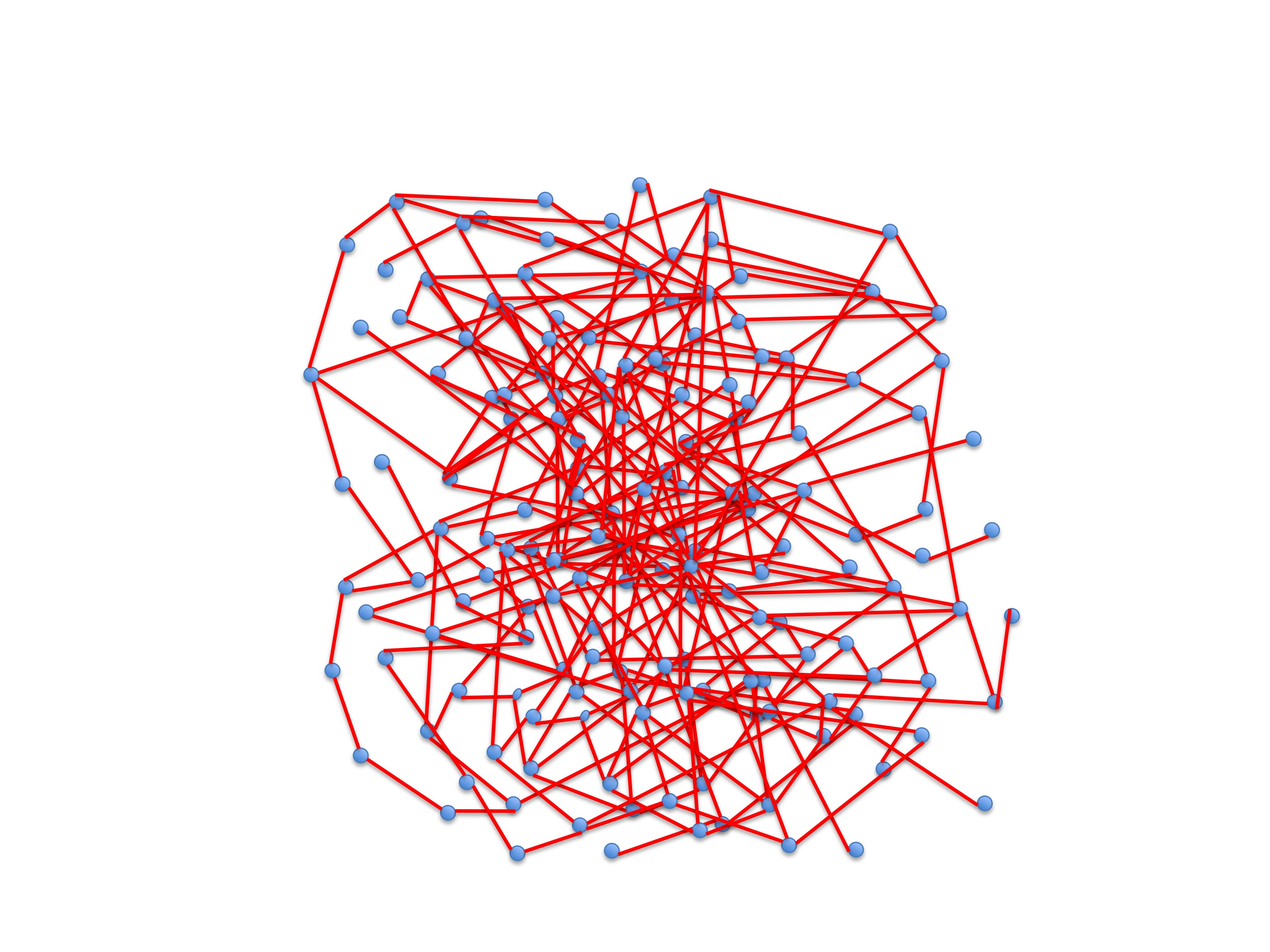}
	\caption{A general Jungle model \label{overflow}}
\end{figure}

Section 4 presents the Jungle model and shows the Jungle model is the optimal probability distribution for modelling losses in a general credit portfolio, under two assumptions:

\begin{enumerate}
	\item The Maximum Entropy principle (to be described in Section 3) is the right guiding principle to select the probability distribution of losses in the framework of credit risk modelling
	\item All the empirical information of a given credit portfolio can be summarized as probabilities of default and default correlations of its constituents
\end{enumerate}

In Section 4, we restrict to (close to) homogeneous portfolios, binomial default indicators and no modelling of recovery rates, without loss of generality as discussed in Section 7.

Section 5 tries to motivate the use of the Jungle model. In particular, we show that when there is no empirical information available on default correlations, the Jungle model becomes the binomial distribution (as it should). We also introduce contagion perturbatively around the binomial model, and we show the behaviour of the corresponding interacting system is the one we would expect intuitively.

Section 5 continues with the presentation of a two particular cases of the Jungle model, the Dandelion model and the Diamond model, both of them are interacting models through contagion.

\begin{figure}[!ht]
	\centering	
	\includegraphics[width=7cm]{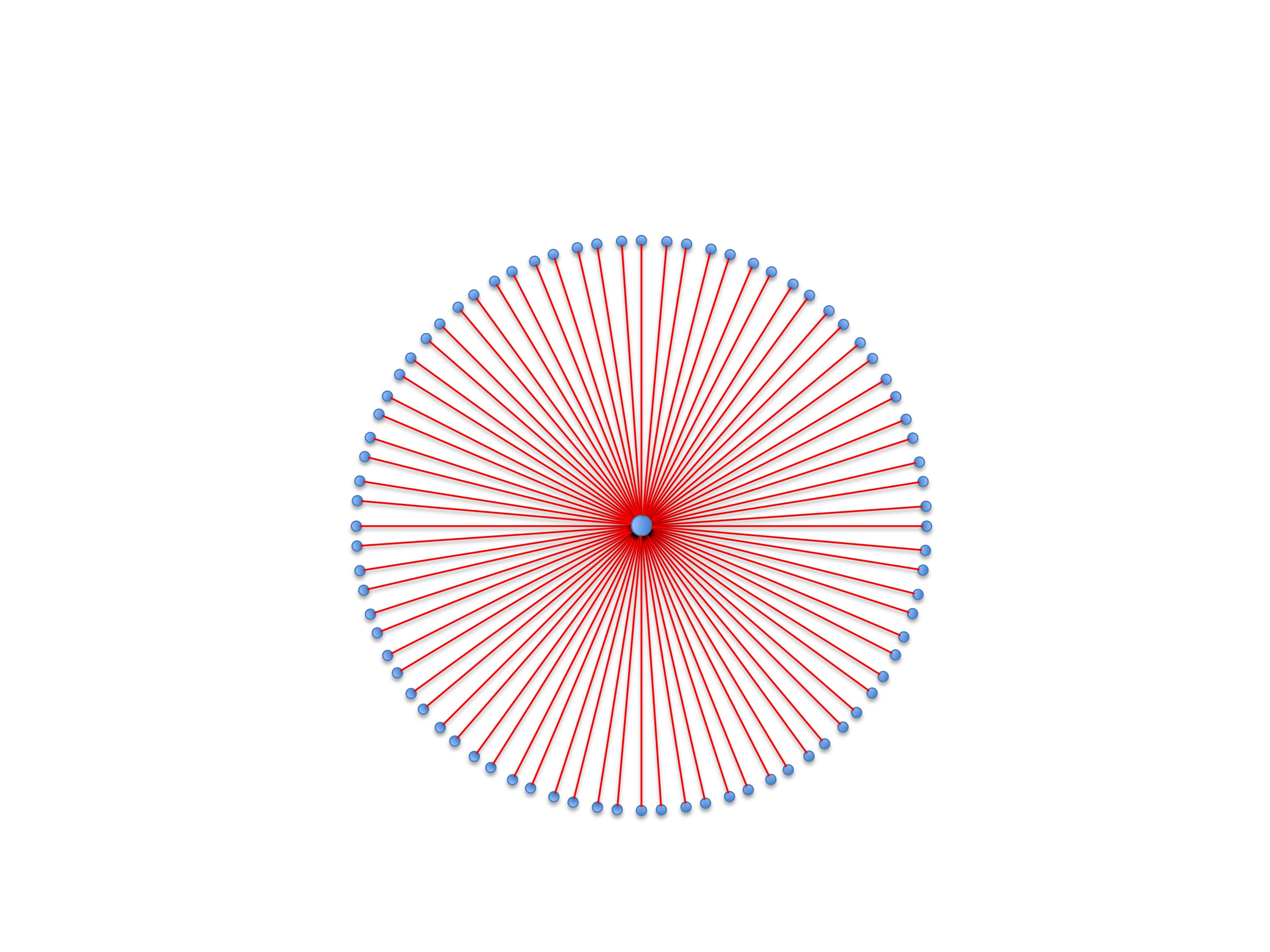}%
	\qquad
	\includegraphics[width=7cm]{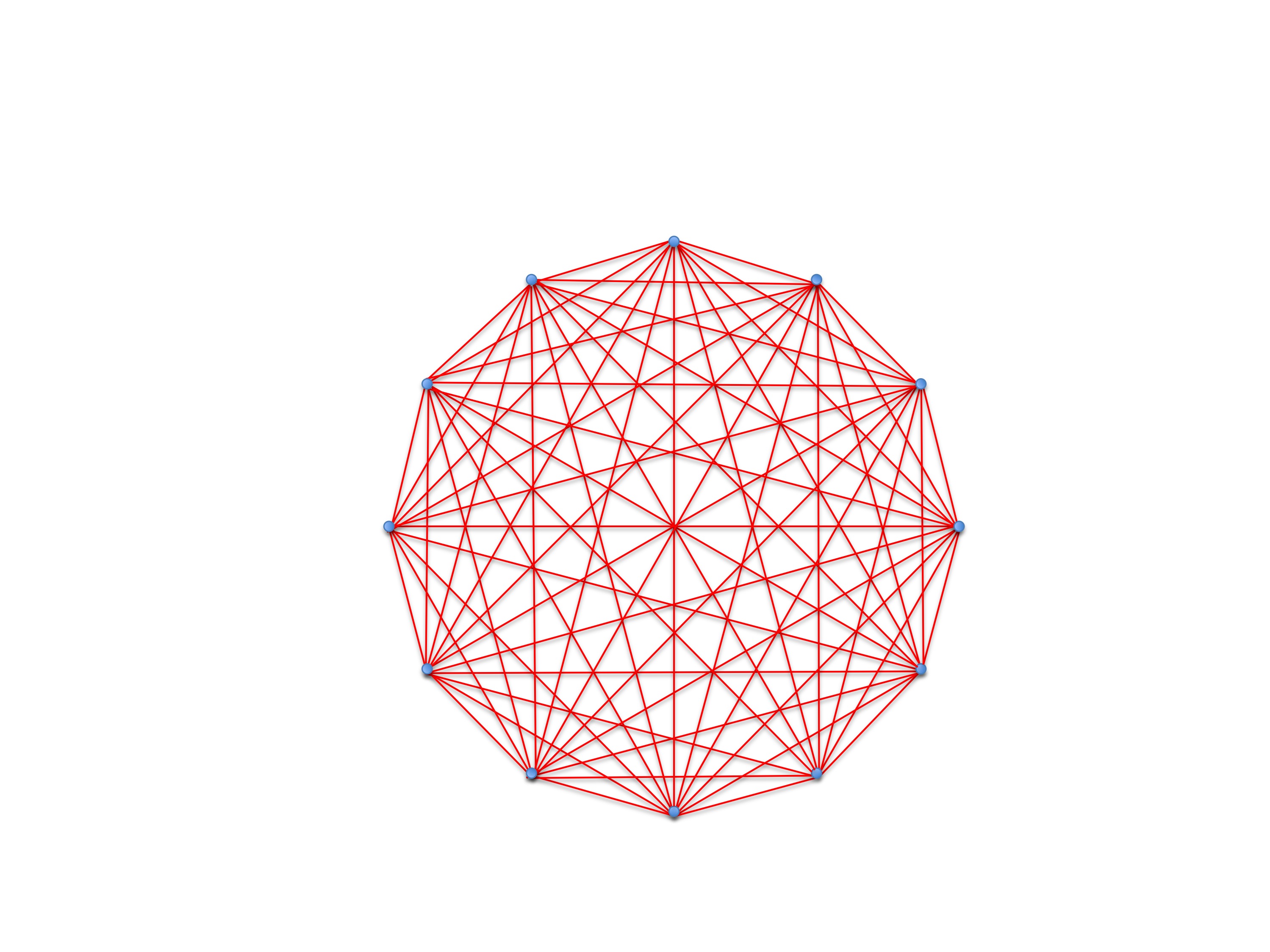}%
	\qquad		
	\caption{A Dandelion and a Diamond}
\end{figure}

The Dandelion model assumes a central element in the credit portfolio is connected through contagion with the rest of the nodes in the portfolio, and no other pair of nodes is connected. Intuitively, the Dandelion model mimics the relationship between a bank and its many borrowers, or even between a Central Bank with the rest of the economy, see \citep{bundes11}.

We show the Dandelion model displays a doubly-peaked loss distribution, endogenously generated through contagion. We also find the results of this model can be interpreted as an endogenously generated two-valued mixture model: the two states of the central node can be understood as the two states of the economy, with the probability of default in the "bad" state of the economy being higher than the probability of default in the "good" state of the economy, by an amount given by the variable representing contagion. In a sense, the Dandelion model provides a unifying way to think about both contagion and macroeconomic risk factors.

We argue the Diamond model experiences a quasi phase transition for a not unreasonable set of empirical parameters, showing quantitatively that a small change in the empirical data may result in significant changes for the profile of the probability loss distribution, leading to severe systemic risks. There is a pictorial analogy with the phase transition of water into steam: if we increase one degree Celsius the temperature of water at $98$ degrees Celsius, the resulting water at $99$ degrees Celsius continues "being water" (small details will change, for example a thermometer inside the water will show a small increase in its readings, but water will remain "being water"). However, when the temperature increases a further degree Celsius, there is a sudden change in the collective behaviour of water, becoming steam. In an analogous way, when the default correlation in the Diamond model is increased a bit above the "quasi phase transition point's default correlation" (to be calculated from the model), the shape of the probability loss distribution (basically the same for all default correlations below that default correlation) changes to a qualitatively different one (which remains basically the same for all default correlations above that default correlation).

In Section 6, we motivate the use of the Jungle model to study model risk in a credit risk framework, i.e. we show how and when the Jungle model can cope with the inherent uncertainty under systemic credit events, such as the ones presented in the previous section (doubly peaked distributions and quasi phase transitions).

In Section 7, we show the Jungle model can be used to model inhomogeneous portfolios, generalizing straightforwardly the binomial loss indicators in Section 4. Even more, we show the Jungle model can straightforwardly be generalized to cope with state-dependent recovery rates, modelling the stylized fact that recovery rates go down when default rates increase.

In Section 8, we argue the three known sources of default clustering (contagion, macroeconomic risk factors and frailty; see the next subsection, "Related literature", for further details) can be understood under the unifying framework of contagion.

In Section 9, we provide a series of policy implications arising from our contagion models. In particular, we show both the U.S. subprime and the European peripheral crises can be understood as particular instances of quasi phase transitions. Also, we are able to understand qualitatively other empirical evidence, such as the thick tails in the historical probability distributions of credit losses presented in Section 2, as well as the surprising fact that quite often, the worst quality credit portfolios end up with default rates lower than the corresponding ones with a better rating. We also pictorially analyse the "Too Big To Fail" phenomenon under our framework based on contagion, and we compare systemic risks out of contagion for a "financial economy of big corporates" versus an economy of "industrial entrepreneurs".

The final section concludes with a summary of the results.
 
\subsection{Related literature}\label{class}

Recent literature suggest there are three main sources of credit clustering: macroeconomic risk factors, contagion and frailty.

Macroeconomic risk factors, such as S\&P 500 returns or short term rates, are common to all credits in the portfolio. When the economy grows strongly, the conditional probabilities of default are low. On the contrary, when the economy weakens, the conditional probabilities of default increase. The passage in time of the business cycle induces in a natural way a correlation among credits. Many standard credit portfolio models can be understood as particular instances of a mixed binomial model, see \citep{embrechts}.

\citep{azizpour14} and \citep{das07} reject the hypothesis that macroeconomic risk factors are able to fully explain the clustering of corporate defaults by themselves, even though \citep{lando10} argues on the contrary.

Contagion can be understood as direct links among credits, such as the ones in a supply chain, or the bank-creditor relationship. A financial crisis may be a prototypical case of contagion, since banks tend to be highly connected with large parts of the economy, and their financial failure may create a deleveraging, impacting directly on the balance sheet of their borrowers. Contagion was analysed with a dynamical approach in \citep{davis01}, \citep{jarrow01}, \citep{giesecke04}, \citep{schon04}, \citep{lut09}, \citep{steinbacher13} and in an Ising setting by \citep{molins05}, followed by \citep{kitsukawa06} and \citep{filiz2012}.

Frailty can be described as the "Enron effect": once the disputable accounting practices were revealed to the public, the probabilities of default of many other companies, in different sectors and regions, readjusted according to the new information. Most likely, no direct links out of contagion between Enron and those companies ever existed, but default correlations arose nonetheless.

\citep{azizpour14}, \citep{duffie09}, \citep{lando10} and \citep{koopman11} include frailty, contagion or both in order to try and explain the clustering of corporate defaults, on top of macroeconomic risk factors. \citep{azizpour14} and \citep{koopman11} conclude both frailty and contagion are necessary to fully explain the clustering of corporate defaults in their datasets, on top of the macroeconomic risk factors.

This paper differentiates from the rest of the literature on contagion networks in the credit arena by arguing its results are independent from the specific details of the "microscopic" credit interactions. In particular, the Maximum Entropy principle argues that given a set of empirical moments of an, in principle, unknown probability distribution, the "best" probability distribution is the Jungle model. The relevant empirical data in the credit arena is known by market participants to be probabilities of default and default correlations. We show that assuming both the probabilities of default and the default correlations correspond to the empirical moments of the unknown probability distribution of credit losses, the Jungle model arises "naturally" and without the need to impose the specific knowledge of the credit interactions among the constituents of the considered credit portfolio.

\section{The data}\label{class}

We use Moody's All rated Annual Issuer-Weighted Corporate Default Rates, from 1920 to 2010, see \citep{moodys}, and \citep{giesecke11} value-weighted default rates on bonds issued by U.S. domestic nonfinancial firms from 1866 to 2008.

As often discussed in the literature and among practitioners, default rate data tends to have issues regarding its interpretation as default losses. This is even more the case for such a long term data set as the ones we use. Our approach is pragmatic: \citep{moodys} and \citep{giesecke11} data are no-nonsense, since even though the data definition process is probably not "rigorous" enough (and it cannot be), the data probably is robust enough (the data contains several full business cycles in both cases).

One of the reasons we use a longer data set than is customary in the literature (\citep{azizpour14} uses data starting on 1970; \citep{das07} on 1979;  \citep{lando10}, on 1982;  \citep{duffie09}, on 1979) is that our models do not require the use of macroeconomic or firm-specific data, so we can go backwards as far as we want, while there is still default rate empirical data. On the contrary, for example the S\&P 500 was launched on 1957, so a researcher needs a lot of ingenuity to be able to find the corresponding macroeconomic and firm-specific data corresponding to several decades ago.

From \citep{moodys}, using the default rate for speculative grade bonds, as well as the number of defaults corresponding to speculative grade bonds, we are able to compute approximately the total amount of speculative grade bonds for each year. For the rest of the paper, when we try to model speculative grade bonds, we will use the average number, close to 800. 

Unfortunately, the corresponding data for single ratings is not provided in the paper. As a consequence, when we deal with Caa-C ratings, we arbitrarily reduce the number for speculative grade bonds by one order of magnitude, 80.

\section{Credit portfolio modelling}

A credit portfolio consists of N credit instruments. A credit portfolio model is a theoretical construct providing as an output the unconditional probability distribution for the losses of a given credit portfolio (the unknown we will focus our attention on for the rest of the paper).

Moody's KMV \citep{merton}, CreditMetrics \citep{cm}, CreditRisk+ \citep{cr} and CreditPortfolioView \citep{wilson} are commercially available credit portfolio models. Additionally, the Gaussian copula \citep{Li} became a widespread tool to value credit derivatives. None of these credit portfolio models were able to model tail risks adequately during either the U.S. subprime crisis or the sovereign and banking crisis in peripheral Europe. Especially, some observers believe the Gaussian copula was a "recipe for disaster", see \citep{wired}, \citep{ft}, when modelling credit tail risks.

The losses of a credit portfolio can be calculated as:

\begin{equation}
L = \sum\limits_{i=1}^N{L_{i}} = \sum\limits_{i=1}^N{E_{i} (1-RR_{i}) l_{i}}
\end{equation}

where $E_{i}$ denotes the Exposure at Default, i.e. the maximum potential loss out of the credit instrument $i$ (usually, the nominal of the bond or loan), $(1-RR_{i})$ denotes the Loss Given Default (RR stands for Recovery Rate) describing the fraction of the Exposure at Default that is effectively lost when the $i$-th borrower defaults, and $l_{i}$ is an indicator taking values in $\{0,1\}$, and which describes if the $i$-th borrower is defaulted or not.

In general, real world cases, $l_{i}$ variables are stochastic, as well as the recovery rates, and the portfolio is inhomogeneous (in general, $E_{i} \neq E_{j}$ for at last some $i \neq j$). The modelling for the related probability distribution of losses is challenging.

We will state our credit portfolio model has been solved when we have found the probability distribution for the losses of that portfolio, $L$. Our target for the rest of the paper will be to motivate, calculate and analyse the probability distribution of $L$.

From now on and until Section 7, we will make the simplification of analysing homogeneous portfolios (with Exposure at Default set at 1), and we will not model Recovery Rates (which is analogous to assume the Recovery Rates are constant and the same for all borrowers). In Section 7, we will deal with the general case of inhomogeneous portfolios and state-dependent Recovery Rates. We will show the simplifications described above do not represent a loss of generality. As a consequence, the state space simplifies to a set of discrete variables taking values 0 or 1, $\Omega = \{(l_{1},l_{2},\ldots,l_{N}) \mid l_{i} \in \{0,1\}, i = 1, 2, \ldots, N\}$. The loss simplifies to $\ell = \sum\limits_{i=1}^N{l_{i}}$.

The probability distribution of a random variable is, in general, unknown and unobservable \textit{per se}. One possible way to derive it is to aggregate the dynamical, "microscopic" processes underlying the random variable. For example, modern physics has been successful at stating microscopic dynamical laws from first principles (quantum mechanics and quantum field theory), and finding the related macroscopic equations (thermodynamics) through an averaging process called Statistical Mechanics.

However, in social sciences this process is fraught with difficulties. In general, the underlying dynamical processes are unknown. The usual methodology then is as follows:

The probability distribution of a random variable is not an observable. But there are observables of the random variable which can be understood as direct calculations from using the probability distribution. For example, the expected value of the random variable is the first moment of the corresponding probability distribution. The variance and the correlation correspond to the second moments of the probability distribution. Skewness and kurtosis, which are widely used empirical observables, are the third and fourth moments of the distribution.

A mathematically well behaved probability distribution can be fully described by its moments. In particular, our underlying random variable, the loss in a credit portfolio, is a bound variable, so we are not concerned about the possibility of moments becoming infinite when the tail of an unbound probability distribution is "fat enough", see \citep{bouchaud03}. As a consequence, it makes sense to assume that despite the probability distribution being not observable, an analyst may finally recover it through the empirical knowledge of its moments (or in general, through the knowledge of the expected value of a general function; the moments are expected values of polynomials).

The question is then: given the knowledge of all or some of its moments, is there a way to find the general form of the probability distribution of the underlying random variables?

The Maximum Entropy principle, or Maxent, provides a specific answer to this question. Maxent asserts:

Given a finite state space $\Omega$, the probability distribution in $\Omega$ that maximizes the entropy and satisfies the following $m < \mathbf{card}(\Omega)$ constraints, given $m$ different functions in $\Omega$, $f_{k}(x)$, and $m$ fixed numbers $F_{k}$:

\begin{equation}
\langle f_{k}(x) \rangle := \sum\limits_{x \in \Omega} P(x) f_{k}(x) = F_{k}, \quad k=1,2,\cdots,m
\end{equation}

as well as a normalization condition:

\begin{equation}
\langle 1 \rangle := \sum\limits_{x \in \Omega} P(x) = 1
\end{equation}

is:

\begin{equation}
P(x) = \dfrac{1}{Z(\lambda_{1}, \lambda_{2}, \cdots, \lambda_{m})} \exp{\left(- \sum\limits_{i = 1}^m \lambda_{i} f_{i}(x)\right)}
\end{equation}

where $Z$ is called the partition function:

\begin{equation}
Z(\lambda_{1}, \lambda_{2}, \cdots, \lambda_{m}) = \sum\limits_{\Omega} \exp{\left(- \sum\limits_{i = 1}^m \lambda_{i} f_{i}(x)\right)}
\end{equation}

The Lagrange multipliers $\lambda_{i}$ are found by inverting the following set of $m$ equations:

\begin{equation}
F_{k} = \langle f_{k}(x) \rangle = - \frac{\partial \, \log Z(\lambda_{1}, \lambda_{2}, \cdots, \lambda_{m})}{\partial \, \lambda_{k}} , \quad k=1,2,\cdots,m
\end{equation}
 
The intuition behind Maxent is $P(x)$ is the "best" \footnote{We leave the "best" concept undefined} probability distribution an analyst can come up with, assuming all the empirical evidence about the problem at hand is summarized as expected values of functions (the $F_{k}$ numbers and the $f_{k}(x)$ functions, respectively). The expected values are taken over the (unknown) probability distribution $P(x)$. The claim above is further discussed at Appendix A.

It often happens that while the "real" probability distribution of a given system is unknown, some constraints are naturally known. For example, in the trivial case of throwing a dice, we know that whatever the correct probability distribution is, the probabilities for each state (each of the six faces of the dice) must add up to one. In fact, Maxent for the dice gives a uniform probability distribution, with a $p = \frac{1}{6}$ for each of the faces of the dice.

In the same way, if we know, in addition to the fact that all probabilities must add up to one, the expected value of the random variable, Maxent produces the binomial distribution. We will see below that when, in addition to the fact that all probabilities must add up to one, both the expected value of the random variable and its correlations are known, Maxent gives the Jungle model.

Maxent is a general principle which pervades science, see \citep{jaynes}. As a consequence, we feel comfortable enough by stating that Maxent is a reasonable principle to pick the probability distribution of losses for a credit portfolio, consistent with the available empirical data.

Let us apply Maxent to a given credit portfolio:

Our state space is $\Omega = \{(l_{1},l_{2},\ldots,l_{N}) \mid l_{i} \in \{0,1\}, i = 1, 2, \ldots, N\}$, a set of discrete variables, $l_{i}$, taking 0 or 1 values, and representing the default / non-default state of the $i$-th credit. As a consequence, the potential moments we might derive from the (unknown) probability distribution of losses are:

\begin{itemize}
	\item The first order moment, $\langle l_{i} \rangle$. This is the so called probability of default of the $i$-th borrower, $p_{i}$
	\item The second order moment, $\langle l_{i} l_{j} \rangle$, for $i \neq j$. This is directly related to the so called default correlation between the $i$-th and $j$-th borrower, 
	
	\begin{equation}
		\rho_{ij} = \dfrac{q_{ij} - p_{i} p_{j}}{\sqrt{p_{i} (1-p_{i})} \sqrt{p_{j} (1-p_{j})}}, \quad q_{ij} := \langle l_{i} l_{j} \rangle
	\end{equation}
	
	\item The second order moment, $\langle l_{i} l_{j} \rangle = \langle l_{i}^2 \rangle $, for $i = j$. However, since $l_{i}$ only takes values in $\{0,1\}$, it is true that $l_{i}^2 = l_{i}$, so the knowledge of this second moment becomes irrelevant
	\item In general, any power of $l_{i}$, $l_{i}^k$, with $k$ being a natural number, becomes $l_{i}$
	\item The third order moment, $\langle l_{i} l_{j} l_{k} \rangle$, for $i \neq j \neq k$, would correspond to the effect on the creditworthiness of the $i$-th borrower, assuming both the $j$-th and $k$-th borrower also default. This effect is conceivable in theory. However, and as far as we know, there is no serious discussion of this phenomenon in the credit literature.
	\item Any moment of order higher than three is bound to the same discussion as the one for the third order moment above
\end{itemize}

There is a general consensus among practitioners that the corresponding available empirical information for a credit portfolio can be summarized as:

\begin{itemize}
	\item The probability of default of a borrower can generally be estimated, either from CDS for liquid names, or from Internal Ratings models for illiquid bonds or loans. Estimates tend not to be too noisy
	\item The default correlation between two borrowers is harder to estimate that the corresponding probabilities of default. There are no financial instruments similar to the CDS to imply the default correlation, or if there are, they tend to be illiquid and over-the-counter (opaque information), and providing noisy estimates. Having said that, and despite the practical difficulties for its estimation, the consensus is the default correlation exists, it can at least be measured in some cases, and it is a key variable to understand default clustering
	\item Third, and higher, order moments bear no specific names in the credit arena
	\end{itemize}

As a consequence, we claim that (at least in our mental framework, which consists of disregarding dynamical, "from-first-principle" equations, and only considering probability distributions arising from imposing empirical constraints to Maxent) the empirical available information for credit portfolios can be summarized in the probabilities of default and default correlations of its constituents.

Maxent selects the Jungle model as its preferred probability distribution for credit losses, consistent with the available empirical data, as seen in the next section.

\section{The Jungle model and credit risk}\label{class}

We consider a credit portfolio of $N$ credit instruments, with a space state $\Omega = \{(l_{1},l_{2},\ldots,l_{N}) \mid l_{i} \in \{0,1\}, i = 1, 2, \ldots, N\}$.

We consider the set "labeling" the $N$ nodes, $\Theta = \{1,2,\ldots,N\}$ and the set "labelling" the $\dfrac{N (N-1)}{2}$ pairs of nodes, $\varPhi = \{(i,j) \mid i = 1,2,\ldots,N \quad \& \quad j > i \}$ (the pair $ij$ and the pair $ji$ are considered to be the same), and two subsets of those, $\theta \in \Theta$ and $\phi \in \varPhi$.

In consistency with the previous section, we assume the full available empirical information of the corresponding credit portfolio can be summarized as the probabilities of default and the default correlations of its constituents.

We will always consider $\theta = \Theta$, or in other words, we assume it is possible to give estimates of the probabilities of default for all the constituents in the portfolio, but $\phi$ will usually be a proper subset of $\varPhi$, meaning some of, but not all, the default probabilities can be estimated. The general case will be one in which $1 \ll \mathbf{card}(\phi) \ll \dfrac{N (N-1)}{2}$.

Using the framework of Maxent, we claim that given the following empirical data, consisting of default probabilities and default correlations:

\begin{itemize}
\item $p_{i}, \forall i \in \theta$, with $p_{i} \in [0,1]$

\item $\rho_{ij}, \forall (i,j) \in \phi$, with $\rho_{ij} \in [-1,1]$; we define $q_{ij}$ such that the relationship $\dfrac{q_{ij} - p_{i} p_{j}}{\sqrt{p_{i} (1-p_{i})} \sqrt{p_{j} (1-p_{j}) }} = \rho_{ij}, \forall (i,j) \in \phi$ holds
\end{itemize}

Leading to the following empirical constraints:

\begin{itemize}
	\item $p_{i} = \langle l_{i} \rangle$, $\forall i \in \theta$
	
	\item $q_{ij} = \langle l_{i} l_{j} \rangle, \forall (i,j) \in \phi$
\end{itemize}

Maxent picks the Jungle model among all the probability distributions consistent with those constraints \footnote{In the physics literature, the Jungle model is called the Ising model with external field, with both space dependent external fields and space dependent local interactions}:

\begin{equation}
P(l_{1},l_{2},\ldots,l_{N}) = \dfrac{1}{Z} \exp{ \left(\sum\limits_{i \in \theta}{\alpha_{i} l_{i}} + \sum\limits_{(i,j) \in \phi}{\beta_{ij} l_{i} l_{j}}\right)}
\end{equation}

where

\begin{equation}
Z = \sum_{\Omega} \exp{ \left(\sum\limits_{i \in \theta} {\alpha_{i} l_{i}} + \sum\limits_{(i,j) \in \phi}{\beta_{ij} l_{i} l_{j}}\right)}
\end{equation}

The unknown parameters $\alpha_{i}$ and $\beta_{ij}$ have to be found by forcing the probability distribution gives the right estimates for the empirical information at our disposal, i.e. the following constraints are satisfied:

\begin{equation}
p_{i} = \langle l_{i} \rangle = \dfrac{\partial \log Z}{\partial \alpha_{i}}
\end{equation}

\begin{equation}
q_{ij} = \langle l_{i} l_{j} \rangle = \dfrac{\partial \log Z}{\partial \beta_{ij}}
\end{equation}

\section{The Jungle model, hands on}\label{class}

After showing the Maxent principle picks the Jungle model as the probability distribution of choice to analyse credit risk (assuming the full empirical information of the credit portfolio can be summarized as the probabilities of default and the default correlations of its constituents), we try and motivate the Jungle model, by studying some particular instances of the general model.

To accomplish that goal, this section presents a few particular cases of the general Jungle model: the binomial model, adding small contagion to the binomial model, the Dandelion model and the Diamond model.

As discussed in Section 3, we will state a probabilistic credit model has been solved once its probability distribution has been computed, either analytically or numerically. There are two ways to solve a model:

\begin{itemize}
	\item $Z$, the partition function, has been summed analytically. We will show the explicit calculation of $Z$ for the Dandelion model below. We will use this methodology for the rest of Section 5.
	\item If the partition function cannot be summed analytically, Markov Chain Monte Carlo methods may allow to generate realisations of the underlying probability distribution, $P$, without having to know its explicit form. With those realisations, all kind of averages of the distribution can be computed. We will apply this methodology in Section 7, when dealing with inhomogeneous portfolios and state-dependent recovery rates.
\end{itemize}

In this section, we will show the Jungle model is able to introduce credit contagion in a way the standard credit portfolio models cannot. In particular, the Jungle model will be shown to model credit correlations under "normal economic conditions" (in a similar way to a Gaussian copula, at least at a non-quantitative level of discourse), but also to endogenously generate quasi-phase transitions, which can be understood as modelling systemic credit crises, arising "out of nowhere", a phenomenon that by definition, a Gaussian copula cannot cope with (in other words, a model under a Gaussian copula never suffers from systemic crises).

\subsection{The binomial model}\label{class}

For a credit portfolio whose probabilities of default are known and equal to each other, $\{p_{i} := p \mid i = 1, 2, \ldots, N\}$ but whose default correlations $\{\rho_{ij} \mid i = 1, 2, \ldots, N \ \& \  j \neq i\}$ are unknown, the probability distribution chosen by the Maximum Entropy principle is:

\begin{equation}
P(l_{1},l_{2},\cdots,l_{N}) = \dfrac{1}{Z} \exp{\left(\alpha \sum\limits_{i=1}^N{ l_{i}}\right)}
\end{equation}

Due to homogeneity, the distribution above becomes the binomial distribution:

\begin{equation}
P\left(\sum\limits_{i=1}^N{ l_{i}} = \ell \right) =  \binom{N}{\ell} p^\ell (1-p)^{N-\ell}
\end{equation}

with the identification $p = \frac{1}{1 + e^{- \alpha}}$. In other words, for the uncorrelated portfolio, the parameter $\alpha$ can be interpreted as (a simple function of) the probability of default. The proof of this result is given in Appendix B.

Since the binomial distribution corresponds to independent defaults, it makes intuitive sense the Maximum Entropy principle selects it when there is no information whatsoever on empirical correlations.

\subsection{Small contagion}\label{class}

In the previous subsection, we have seen the Jungle model with $\{\alpha_{i} := \alpha \mid i = 1, 2, \ldots, N\}$ and $\{\beta_{ij} = 0, \, \forall (i, j) \in \phi \}$, becomes the binomial distribution. Then, the probability of default of the credit instruments becomes (a simple function of) $\alpha$.

We might ask ourselves which would be the effect on the portfolio of adding the minimum amount possible of $\beta_{ij}$ (a small $\beta$ to only a pair of nodes, say $12$, and $\beta_{ij} = 0$ for any other pair of nodes $ij$ different from $12$) to the Jungle model corresponding to the binomial distribution (only with $\alpha$), or in other words, we are interested in expanding perturbatively around the binomial model, in order to single out the effect of $\beta_{ij}$, i.e. to see if $\beta_{ij}$ can be interpreted in relation to the empirical parameters, $p$ and $\rho$, in the same way for the binomial distribution we could interpret $\alpha$ as (a simple function of) the underlying $p$. 

The corresponding probability distribution for the losses of that portfolio is:

\begin{equation}
P_{\beta}(l_{1},l_{2},\cdots,l_{N}) = \dfrac{1}{Z} \exp{\left(\alpha \sum\limits_{i=1}^N{ l_{i}} + \beta l_{1} l_{2}\right)}
\end{equation}

The answer to our question is $\rho_{12}^{\beta}$ is proportional to $\beta$, for small $\beta$ and for a given probability of default, as it can be seen from Appendix C. In other words, when small amounts of contagion are added to an uncorrelated credit portfolio, the default correlation increases (from 0). And the rate of increase is proportional to $\beta$, so for small contagion, the coefficient $\beta$ can be interpreted as (a simple function of) the default correlation, in the same way that for no contagion, $\alpha$ can be interpreted as (a simple function of) the probability of default.

Also, we can see that $p_{1}^{\beta} = p_{2}^{\beta}$, by symmetry. But $p_{1}^{\beta} > p_{1}^{\beta=0}$, with the increase being proportional to $\beta$. Instead, $p_{j}^{\beta} = p_{j}^{\beta=0}, j=3, \cdots, N$, since the nodes $j=3, \cdots, N$ are not affected by contagion.

In other words, when some contagion is added, it is not true any more that $\alpha$ is (a simple function of) $p$, since there is a "mixing" between $\alpha$ and $\beta$ and their relationships with respect to $p$ and $\rho$.

We want to emphasize that the model above does not correspond to a credit portfolio whose probabilities of default are known and equal to each other, $\{p_{i} =: p \mid i = 1, 2, \ldots, N\}$ and the default correlation is only known for the pair of nodes $12$,  $\beta_{12} =: \beta$ and $\beta_{ij} = 0,$ for $ij \neq 12$. 

As we have seen above, the probabilities of default for the model described by:

\begin{equation}
P_{\beta}(l_{1},l_{2},\cdots,l_{N}) = \dfrac{1}{Z} \exp{\left(\alpha \sum\limits_{i=1}^N{ l_{i}} + \beta l_{1} l_{2}\right)}
\end{equation}

are not the same for all nodes: credit instruments with a contagion link, such as $l_{1}$, experience an increase in their probabilities of default, with respect to those nodes without a contagion link, such as $l_{3}$.

The probability distribution satisfying the empirical conditions such that the probabilities of default are known and equal to each other, $\{p_{i} =: p \mid i = 1, 2, \ldots, N\}$ and the default correlation is only known for the pair of nodes $12$,  $\beta_{12} =: \beta$ and $\beta_{ij} = 0,$ for $ij \neq 12$ is:

\begin{equation}
P_{\beta}(l_{1},l_{2},\cdots,l_{N}) = \dfrac{1}{Z} \exp{\left(\alpha_{0} (l_{1} + l_{2}) + \alpha \sum\limits_{i=3}^N{ l_{i}} + \beta l_{1} l_{2}\right)}
\end{equation}

Where $\alpha_{0}$ is such that the constraint $\langle l_{1} \rangle_{\beta} = \langle l_{2} \rangle_{\beta} = p$ is satisfied, being different from the $\alpha$ required to satisfy the constraint $\langle l_{3} \rangle_{\beta} = \cdots \langle l_{N} \rangle_{\beta} = p$. For this case, it is also true that for small $\beta$, the default correlation of the pair $12$ is proportional to $\beta$.

In the general Jungle case in which the model contains both $\alpha_{i}$ and $\beta_{ij}$:

\begin{equation}
P(l_{1},l_{2},\ldots,l_{N}) = \dfrac{1}{Z} \exp{ \left(\sum\limits_{i \in \theta}{\alpha_{i} l_{i}} + \sum\limits_{(i,j) \in \phi}{\beta_{ij} l_{i} l_{j}}\right)}
\end{equation}

it will not be true any more that $\alpha$ is (a simple function of) the probability of default, and $\beta$ is (a simple function of) the default correlation: for the general Jungle case, there is a "mixing" between $\alpha$ and $\beta$ and their relationships with respect to $p$ and $\rho$.

\subsection{The Dandelion model}\label{class}

The Dandelion model corresponds to a Jungle model with $N+1$ borrowers, such that the first one, defined as $i=0$ and considered to be at the centre of the Dandelion, is "connected" to all remaining borrowers, at the external surface of the Dandelion, such that $\beta_{0i} =: \beta \neq 0$ for $i = 1, 2, \cdots, N$. Any other borrowers remain unconnected, $\beta_{ij} = 0$ for $i = 1, 2, \cdots, N \quad \& \quad j > i$. For simplicity, we assume $\alpha_{i} =: \alpha$ for $i = 1, 2, \cdots, N$.

The probability distribution for the Dandelion model is:

\begin{equation}
P(l_{1},l_{2},\ldots,l_{N}) = \dfrac{1}{Z} \exp{ \left(\alpha_{0} l_{0} + \alpha \sum\limits_{i = 1}^N l_{i} + \beta \sum\limits_{i=1}^N l_{0} l_{i}\right)}
\end{equation}

The Dandelion model, despite being interacting, can be fully solved, with the probability distribution for its losses given by:

\begin{equation}
P(\ell = \sum\limits_{i=1}^N l_{i}) = \dfrac{1}{Z} \binom{N}{\ell} \left(\exp{(\alpha \ell)} + \exp{(\alpha_{0} + \ell (\alpha + \beta))} \right)
\end{equation}

where Z is given by:

\begin{equation}
Z = (1 + e^{\alpha})^N + e^{\alpha_{0}}  (1 + e^{\alpha + \beta})^N
\end{equation}

and $\alpha$, $\alpha_{0}$ and $\beta$ are given explicitly as functions of the empirical data, $p$, $p_{0}$ and $\rho$:

\begin{align}
\alpha_{0} = (N-1) \log\left(\frac{1-p_{0}}{p_{0}}\right) + N \log\left(\frac{p_{0} - q}{1 - p_{0} - p + q}\right)\\
\alpha = \log\left(\frac{p - q}{1 - p_{0} - p + q}\right)\\
\beta = \log\left(\frac{q}{p_{0} - q} \frac{1 - p_{0} - p + q}{p - q}\right)
\end{align}

where $q$ can be derived from the definition of default correlation:

\begin{equation}
\rho = \frac{q - p p_{0}}{\sqrt{p (1 - p)} \sqrt{p_{0} (1 - p_{0})}}
\end{equation}

The proof can be found in Appendix D.

To provide intuition for the Dandelion model, we have calculated its probability distribution for a set of reasonable parameters, $N = 800$ and $p = p_{0} = 2.8\%$, which correspond to the historical default rate average for global speculative-grade bonds, as per \citep{moodys}, and for a given range of possible default correlations. The result can be found in Figure 4:

\begin{figure}[!htbp]
\centering
\includegraphics[width=11cm]{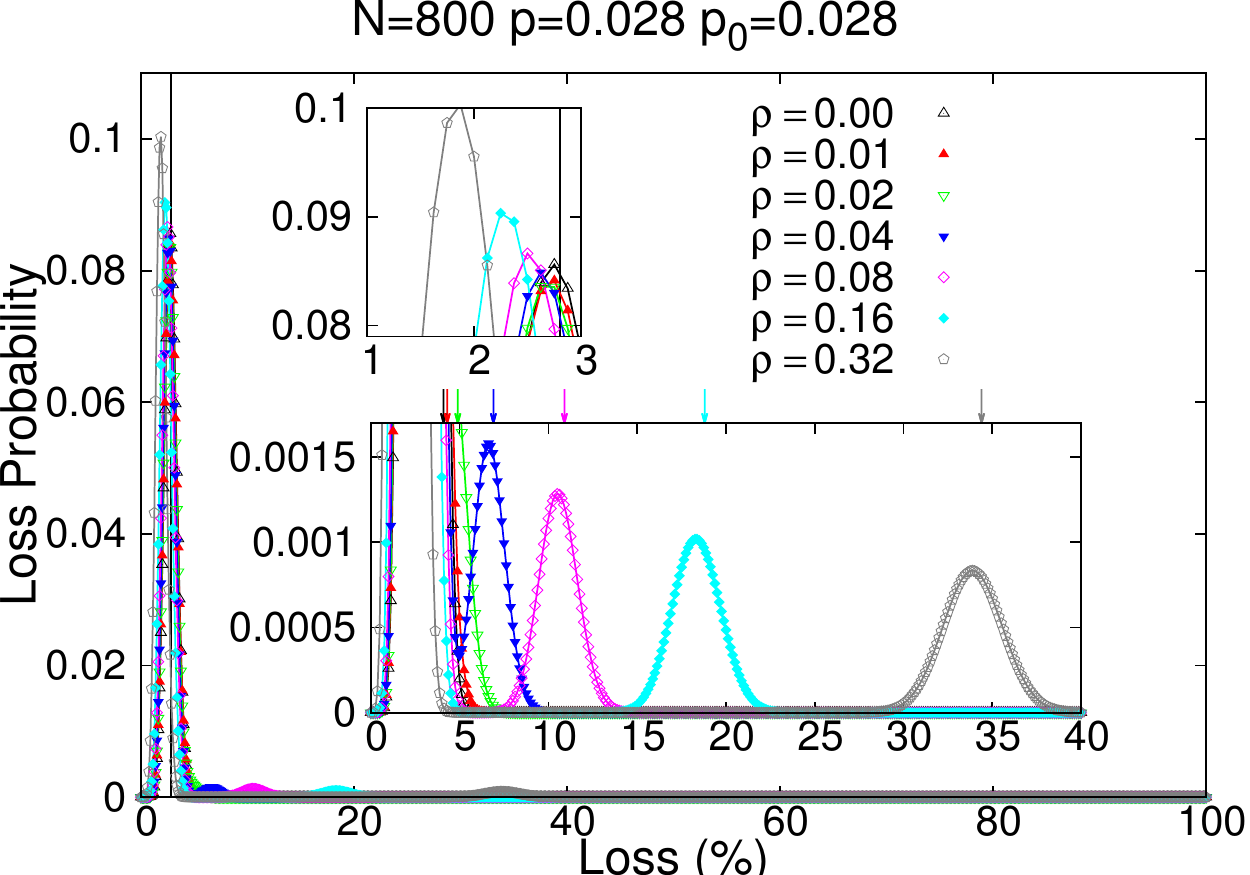}
\caption{Probability distributions for the losses of the Dandelion model, corresponding to different default correlations}
\end{figure}

The probability distributions in the chart show a "double peak" pattern: on one hand, a first peak, centred at low losses and not unlike the corresponding peak for a binomial distribution. On the other hand, a smaller but not insignificant second peak, corresponding to a high level of losses, and consistent with avalanches / domino effects due to contagion.

The higher the default correlation, the higher the extreme losses (the second peak moves further to the right on the chart). Also, the higher the default correlation, the lower losses on the first peak. Contagion works both ways: defaults lead to more defaults (with respect to the binomial case), non-defaults lead to more non-defaults (with respect to the binomial case). These two effects can be seen more specifically from the two insets in the chart.

Also, the higher the correlation of default, the higher the Value at Risk and the Expected Shortfall. The dependency of these two risk measures with respect to the corresponding default correlation is exemplified by the following table (at the $99\%$ confidence level).

\bigskip

\begin{table}[!htbp]
\centering
\begin{tabular}{c|cc}
	$\rho$ & VaR & ES \\
	\hline
	0.00 & 0.041 & 0.044 \\
	0.01 & 0.043 & 0.046 \\
	0.02 & 0.049 & 0.055 \\
	0.04 & 0.069 & 0.076\\
	0.08 & 0.109 & 0.117\\
	0.16 & 0.188 & 0.198\\
	0.32 & 0.344 & 0.356\\
	\hline
\end{tabular}
\end{table}

The Dandelion model can be understood as a bridge between macroeconomic risk factors and contagion. Specifically, in the derivation of the Dandelion model in Appendix D, the following equation arises:

\begin{equation}
p = (1-p_{0}) p(\alpha) + p_{0} p(\alpha + \beta)
\end{equation}

where 

\begin{equation}
p(\alpha) = \frac{1}{1 + e^{- \alpha}}
\end{equation}

corresponds to the relationship between $p$ and $\alpha$ described for the binomial (non-interacting) case.

As a consequence, the central node in the Dandelion could be interpreted as endogenously generating a "macroeconomic state of the economy", whereby for a fraction of time given by $1 - p_{0}$ the economy remains in a "good" state of the economy, with a probability of default for its constituents given by $p(\alpha) = \frac{1}{1 + e^{- \alpha}}$, and for a fraction of time given by $p_{0}$ the economy remains in a "bad" state of the economy, with a probability of default for its constituents given by $p(\alpha +  \beta) = \frac{1}{1 + e^{- (\alpha +  \beta)}}$, where $p(\alpha +  \beta) > p(\alpha)$, and the difference is accounted by the "contagion factor" $\beta$.
	
In other words, the Dandelion model endogenously generates a kind of mixture of binomials, able to generate a doubly peaked distribution and clustering of defaults.

\subsection{The Diamond model}

The Diamond model is defined by:

\begin{equation}
Z = \sum_{l_{1},l_{2},\ldots,l_{N}} \exp{ \left(\alpha \sum\limits_{i=1}^N { l_{i}} + \beta \sum\limits_{i > j}{ l_{i} l_{j}}\right)}
\end{equation}

The Diamond model describes a set of credits, all interacting among each other. For example, if $N = 4$, node 1 could be a bank, node 2 a cement producer, node 3 a real estate developer and node 4, a car dealer. The cement producer, the real estate developer and the car dealer get financing from the bank, so there are correlations of defaults between the pairs 12, 13 and 14. Also, the cement producer is a supplier to the real estate developer, so the pair 23 is also correlated. Finally, workers at firms 2 and 3 purchase cars from the car dealer, so a default of 2 or 3 would impact on 4 business, creating also default correlations between 24 and 34.

The partition function for the Diamond model is given by:

\begin{equation}
Z = \sum_{\ell=0}^N \binom{N}{\ell} \exp{ \left((\alpha - \frac{\beta}{2}) \ell + \frac{\beta}{2} \ell^2\right)}
\end{equation}

And the corresponding probability distribution for the losses will be:

\begin{equation}
P(\ell = \sum\limits_{i=1}^N l_{i}) = \binom{N}{\ell} \frac{ \exp{ \left((\alpha - \frac{\beta}{2}) \ell + \frac{\beta}{2} \ell^2\right)}}{Z}
\end{equation}

We can relate the empirical data, $p$ and $\rho$ to the model parameters $\alpha$ and $\beta$, from the following two equations which can be inverted numerically:

\begin{align}
p = \frac{1}{Z N} \sum_{\ell=0}^N \binom{N}{\ell} \ell \exp{ \left((\alpha - \frac{\beta}{2}) \ell + \frac{\beta}{2} \ell^2\right)}\\
q =\frac{2}{Z N (N-1)} \sum_{\ell=0}^N \binom{N}{\ell} \frac{1}{2} \ell (\ell-1) \exp{ \left((\alpha - \frac{\beta}{2}) \ell + \frac{\beta}{2} \ell^2\right)}
\end{align}

Appendix E gives a proof of the previous statements.

The Diamond model clearly exemplifies one of the most interesting phenomena of the Jungle model: quasi phase transitions. 

Let us see how the probability distribution of losses for the Diamond model changes, when we smoothly change default correlations, for the probability of default fixed at a given level (with parameters $N=20$ and $p = 40\%$, for easiness of visual inspection; below, we will provide another example, with $N=50$ and $p = 2.8\%$):


\begin{figure*}[!ht]
	\centering
	
	\includegraphics[width=15cm]{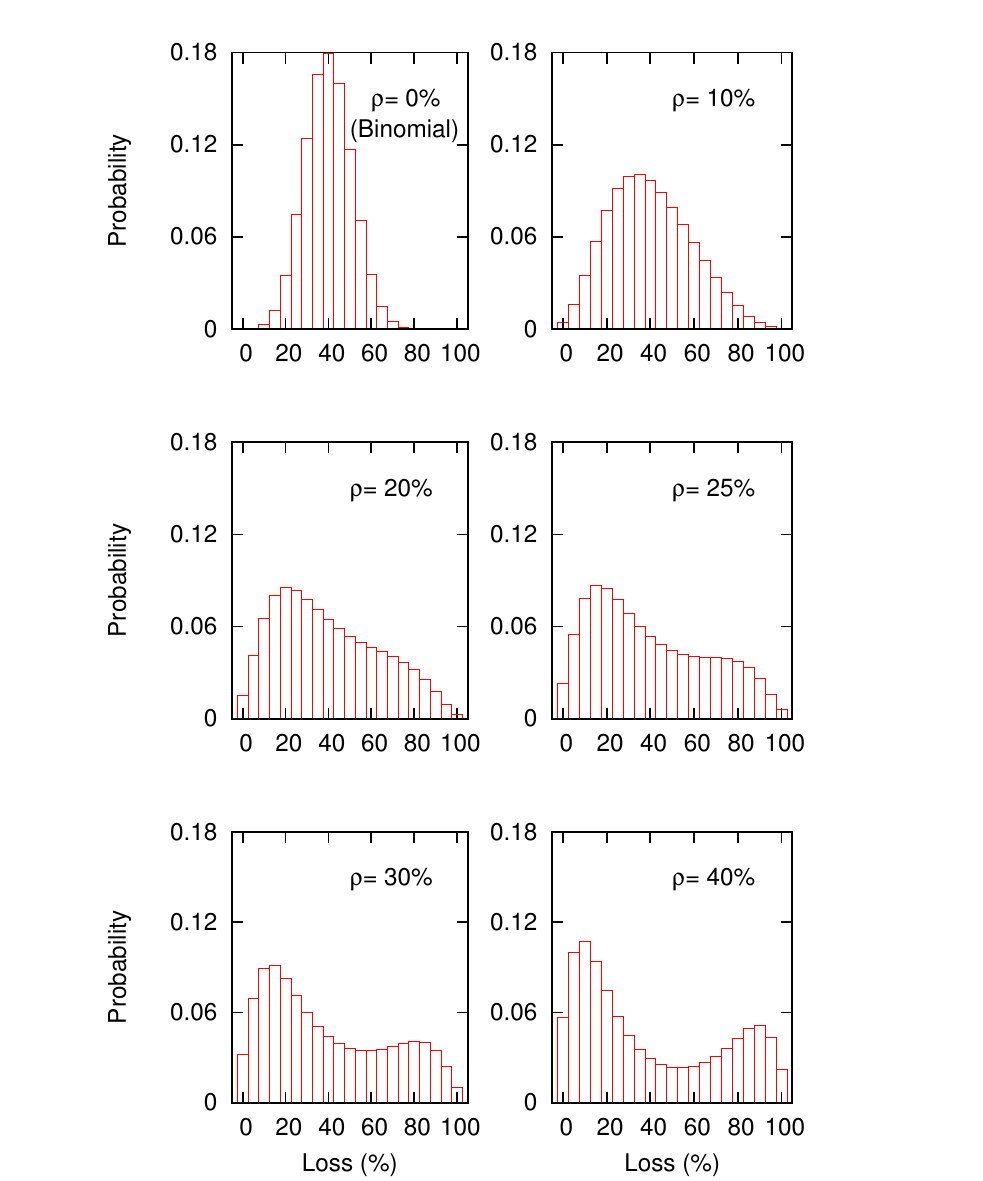}%
	\caption{Loss probability distributions for default correlations below, around and above the quasi phase transition point}
\end{figure*}

We can see there is a sudden change of collective behaviour for the probability distribution of losses when we smoothly change $\rho$ from 10\% to 20\% to 30\%, at some point between these default correlations: 

For default correlations at around 10\% or below, the Diamond model presents a standard behaviour with losses spread with a given width around the expected value, 40\%. However, when the default correlation increases only slightly (to 25\%, say), a different behaviour for the probability distribution of losses starts to emerge: the probability distribution for the losses becomes bimodal, as it can be seen from Figure 5. And the more the default correlation increases, the larger the potential losses out of the second peak on the right.

Another numerical example, this time with $N=50$ and $p = 2.8\%$, the average default rate for speculative-grade bonds in the \citep{moodys} sample, shows how a quasi phase transition changes dramatically the risk profile of the loss probability distribution, given small changes of the empirical values determining the portfolio (probabilities of default and, especially, default correlations):


\begin{figure*}[!ht]
	\centering

	\includegraphics{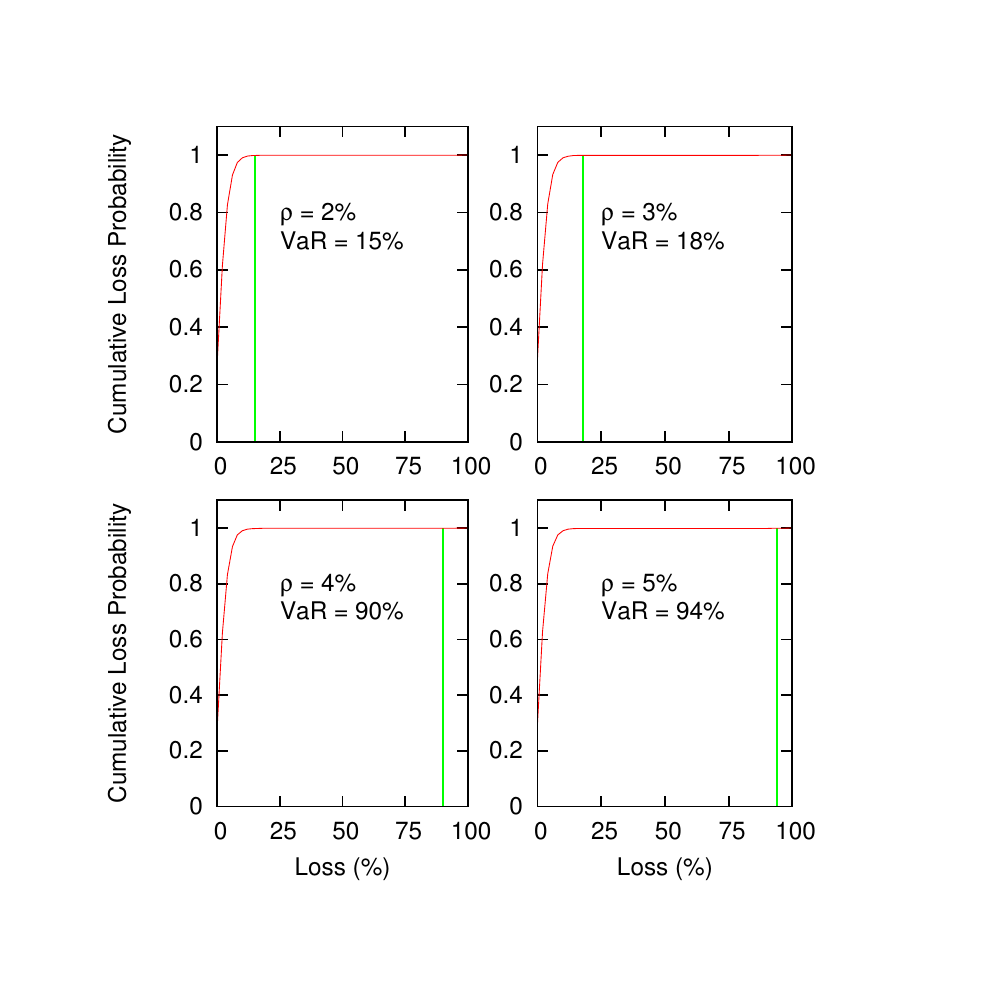}%
	\caption{Cumulative probability loss distributions for default correlations below, around and above the quasi phase transition point}
\end{figure*}

From Figure 6, we can see a sudden jump for the Value at Risk at the $99.9\%$ confidence level, given a small increase in the default correlation \footnote[1]{This is the explanation for the use of the "phase transition" concept, borrowed from Statistical Mechanics. Phase transitions suffer a sudden jump in a given variable, induced by a small change in another, underlying variable. However, a quasi phase transition is not a phase transition, as properly defined in Statistical Mechanics. For example, phase transitions for the Ising model, the equivalent of the Jungle model in Physics, cannot happen for finite $N$, and throughout the paper we assume $N$ is always finite.}.

The above phenomenon can also be analysed by looking at how the empirical variables, both the probability of default and the default correlation, change when the model parameters, $\alpha$ and $\beta$, change, as shown in Figure 7.

Despite the fact that for a finite $N$, there can be no (exactly discontinuous) phase transition, from the figures above we can see a sharp, almost discontinuous behaviour throughout a clearly visible diagonal line in the space of model parameters. By analogy, we give the name "quasi phase transition" to that phenomenon. The existence of such a line, ending in the so called "critical point"\footnote[2]{It is known in the Physics literature the critical point lies at $\alpha = -2$, $\beta=\frac{4}{N}$ in the large $N$ limit, corresponding approximately to a probability of default of $44\%$ and a default correlation of $11\%$ for $N=80$.}, is well known in the Condensed Matter physics literature.

Since there is a known relationship between the empirical parameters, the probability of default and the default correlation, and the model parameters, $\alpha$ and $\beta$, an analyst can use either the empirical parameters or the model parameters to describe the behaviour of the model. The model parameters tend to be more useful to analyse the behaviour of such systems, because as observed from Figure 7, a small change in the empirical parameters always results in a small change for the model parameters, but the opposite is clearly not true (at the line of quasi phase transitions):

\begin{figure*}[!ht]
	\centering
	\includegraphics[width=7cm]{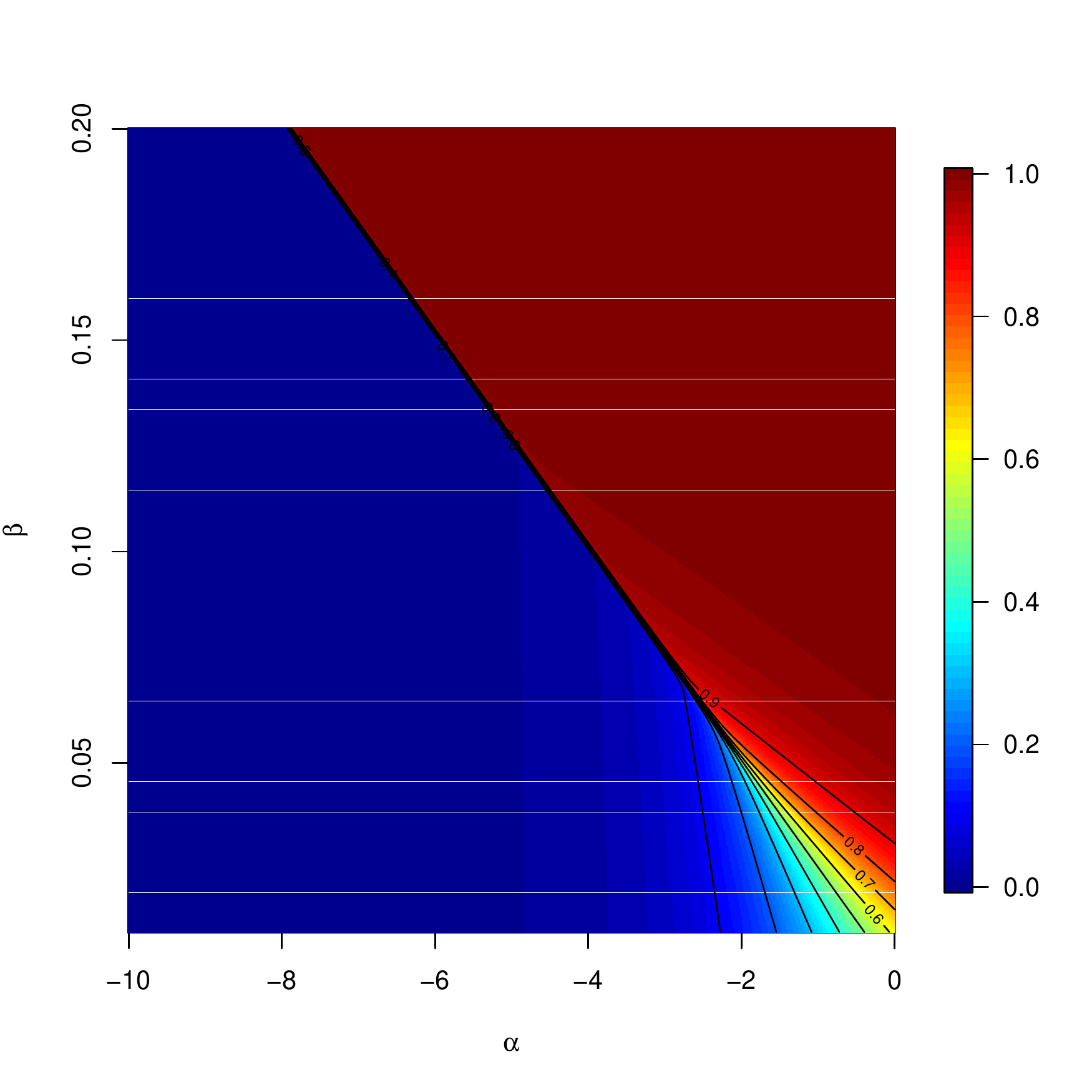}%
	\qquad
	\includegraphics[width=7cm]{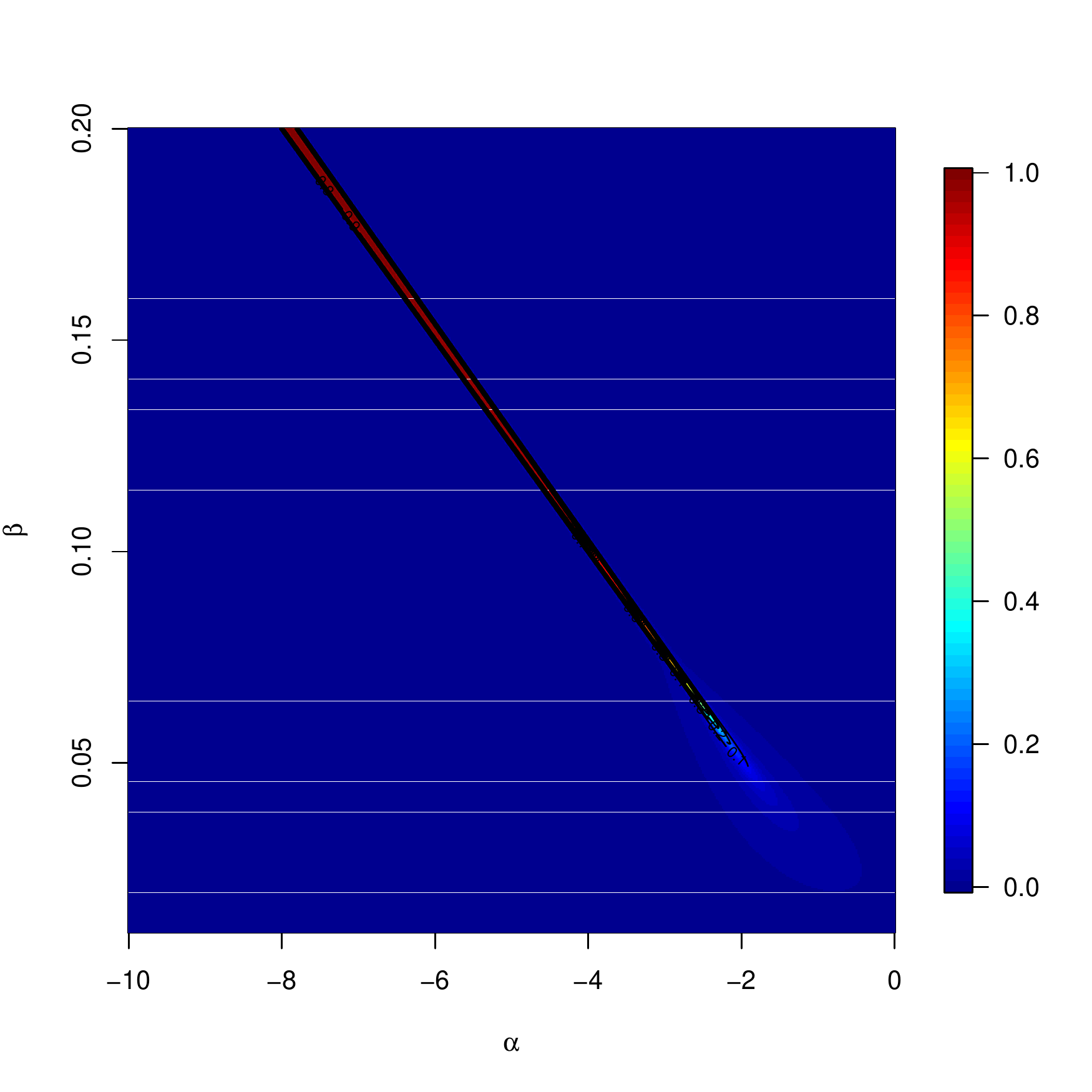}%
	\qquad
	\caption{Probability of default and default correlation for a given set of model (normalized to values between 0 and 1) parameters, $\alpha$ and $\beta$, and $N=80$.}
\end{figure*}

Also, any credit portfolio is driven by underlying, fundamental economic factors (both macroeconomic and microeconomic). As a consequence, we can understand the evolution in time of a credit portfolio described by such a model as the smooth change of the model parameters when the underlying, fundamental economic factors change. Usually, the empirical parameters will be such that the system will move around the bottom left corner of the figure 7 (relatively low probabilities of default and default correlations). 

However, when the economic conditions are such that the system is close to (but still below) the line of quasi phase transitions, a small change in the underlying, fundamental economic factors may lead to a small change in the model parameters, such that the system may inadvertently cross the line of quasi phase transitions, resulting in an abrupt, almost discontinuous, change in the empirical parameters.

As a consequence, the Diamond model shows anti-intuitively that the collective behaviour of the portfolio may significantly change due to small changes of the empirical values determining it.

This phenomena is not unlike the phase transition of water into steam: if we increase one degree Celsius the temperature of water at $98$ degrees Celsius, the resulting water at $99$ degrees Celsius continues "being water" (small details will change, for example a thermometer inside the water will show a small increase in its readings, but water will remain "being water"). However, when the temperature increases a further degree Celsius, there is a sudden change in the collective behaviour of water, becoming a coexistence of steam bubbles and liquid.

So, a small change of the underlying parameters leads to a significant change of the behaviour for the whole system. This is surprising, since if we could solve all the dynamical equations of motion for the say $10^{23}$ particles in a litre of water, it seems unlikely that with that knowledge we could have forecasted such a dramatic change of behaviour. It is the averaging out of "irrelevant" degrees of freedom, undertaken by statistical mechanics, which allows to keep only the (small set of) parameters which really matter at the level of one litre of water.

Analogously, the Diamond model shows a quasi phase transition from a phase dominated by a "binomial-like" behaviour, whereby losses spread over a given width, centred around the expected loss, towards a coexistence region, dominated by avalanches due to credit contagion, and determined by a doubly peaked distribution. The transition from one phase to the coexistence region is caused by a smooth change of the empirical parameters defining the portfolio (probabilities of default and default correlations). However, the variation in the global shape of the probability distribution changes significantly the risk profile of the portfolio, potentially inducing systemic risks.

\subsection{The Jungle model and the real world}\label{class}

In a general, real world credit portfolio, a Jungle model will be defined by its topology, $\theta$ and $\phi$, as well as by the given empirical data, consisting of $p_{i}$ and $\rho_{ij}$ over $\theta$ and $\phi$.

The data will always be such that $\theta = \Theta$, or in other words, we consider it is possible to give estimates of the probabilities of default to all the constituents in the portfolio, but $\phi$ will usually be a proper subset of $\varPhi$, meaning some of, but not all, the default probabilities can be estimated. The general case will be one in which $1 \ll \mathbf{card}(\phi) \ll \dfrac{N (N-1)}{2}$.

Pictorially, the network corresponding to that credit portfolio will be a possibly random combination of links connecting many nodes in the network. But quite often, the analyst will be able to recognize Dandelion shapes (possibly centred at banks or other large corporates) and Diamond shapes, among others. As a consequence, the ability to solve exactly these three interacting models may prove helpful.

\section{The Jungle model and model risk}\label{class}

Above, we have shown the Jungle model becomes the binomial model when no information whatsoever about correlations is known. This result is intuitive, since the binomial model describes losses for independent defaults.

In general, when there is correlation among some of the underlying credit instruments, the Jungle model will naturally depart from the binomial model.

Model risk in the framework of credit portfolio modelling is the risk that a given probability distribution for the losses underestimates the tail risks (with respect to empirical evidence).

Pictorially, we could think of a "functorial" assigning a probability distribution of losses for each theory. For example, the functorial would assign the binomial probability distribution to the "binomial theory", and it would assign the Jungle probability distribution to the "Jungle theory".

We could try to parametrize the space of potential theories. For example, we could assign parameter $1$ to the binomial theory (given that the binomial theory depends on a parameter $p$, directly related to its first moment), we could assign parameter $2$ to the Jungle theory (given that the Jungle model depends on parameters $p$ and $\rho$, directly related to its first and second moments) and so on.

It seems clear that restricting to parameter $1$ (most standard credit models can be understood as straightforward generalizations of the binomial theory) creates a significant model risk. Empirically, this has been seen during the recent financial crises in the Western world.

An analyst could believe modelling credit risk with theories with parameter $2$, i.e. the Jungle model, would be clearly better than restricting to parameter $1$ (since the binomial theory is a particular case of the Jungle model). However, it seems that theories with parameter $2$ are only slightly "larger" than theories with parameter $1$, but the whole space of theories is vastly larger than $2$. We could see ourselves then climbing the ladder, enlarging progressively the space of theories, but always aware that we would suffering from a massive model risk, since the "real theory" could be one with $n > 2$ parameter, whatever $n$ could be.

Empirically, probabilities of default can be obtained easily for a wide range of companies. Default correlation data is not as easily available as probabilities of default are, but it is a relevant parameter for practitioners and academics alike. However, expected values of cross-products of three or more $l_{i}$, such as $\langle l_{i} l_{j} l_{k}\rangle$, for $i \neq j \neq k$, are not known. As far as we know, these terms have not been seriously considered in the related literature. 

As a consequence, probably the Jungle model is not the most general credit risk model. The "correct" credit risk model could be one with $n > 2$ parameter \footnote{If new empirical data were known in the form of higher order moments, the framework in this paper could cope with that. In fact, in Statistical Physics this kind of extended Jungle models, including trios and higher order interactions, have been studied extensively}, and this is unknown to us. Having said that, Maxent picks the Jungle model as its credit risk model of choice, in consistency with the available empirical data (probabilities of default and default correlations).

In other words, the Jungle model is the best we can do with the empirical credit data at our disposal.

Another issue related to model risk is what do we mean by "available empirical data": any sample data is bound to intrinsic uncertainty. Not only because of fluctuations over time, but also due to imperfections in how data is presented and collected (there are hundreds of different day count conventions in finance, for example).

And small fluctuations in the empirical data may have a big impact on the selected model. In the presentation so far, we have not discussed how do we select in practice the parameters in the Jungle model, $\alpha_{i}$ and $\beta_{ij}$, to fit with the empirical data $p_{i}$ and $\rho_{ij}$, apart from stating (as per Maxent) that the constraints:

\begin{equation}
p_{i} = \langle l_{i} \rangle = \dfrac{\partial \log Z}{\partial \alpha_{i}}
\end{equation}

\begin{equation}
q_{ij} = \langle l_{i} l_{j} \rangle = \dfrac{\partial \log Z}{\partial \beta_{ij}}
\end{equation}

have to be satisfied. In the previous section, we were able to invert analytically the relationship of $p_{i}$ and $\rho_{ij}$ with respect to $\alpha_{i}$ and $\beta_{ij}$ for the Dandelion model. For the Diamond model, we showed the corresponding equations can be solved numerically. 

However, for a general Jungle model, the situation is probably much more precarious. We might have a large amount of bonds and loans, $N$, also $N$ probabilities of default, and a large number (much larger than $1$, but much smaller than the maximum possible amount of links, $\frac{N (N-1)}{2}$) of default correlations among the borrowers. And it could perfectly be possible we could not sum the partition function $Z$ analytically, so we would need to resort to MCMC methods.

In this situation, a "Jungle inverter" (i.e., a function providing $\alpha_{i}$ and $\beta_{ij}$ on $\theta$ and $\phi$, given a set of empirical values for $p_{i}$ and $\rho_{ij}$ on $\theta$ and $\phi$, see for example \citep{roudi09}) would be noisy. A good "Jungle inverter" could give the "correct" $\alpha_{i}$ and $\beta_{ij}$ if supplied with the "correct" $p_{i}$ and $\rho_{ij}$. But as discussed above, it will never be possible to provide the "correct" empirical data, we will always be using sample data, prone to an unavoidable margin of error.

We suggest the following way of thinking about model risk: 

The empirical data $p_{i}$ and $\rho_{ij}$ has not to be thought as a point in a $(N + \frac{N (N-1)}{2})$-dimensional space, but as a $(N + \frac{N (N-1)}{2})$-dimensional cube centred at $p_{i}$ and $\rho_{ij}$, with a certain width, $\delta p_{i}$ and $\delta \rho_{ij}$.

We should then sample randomly a point from that cube, $\overline{p_{i}}$ and $\overline{\rho_{ij}}$. The Jungle inverter would give us a set of $\overline{\alpha_{i}}$ and $\overline{\beta_{ij}}$. Sampling again from the cube, we would get another set of parameters, and so on. 

Due to the large scale of the inversion, it seems reasonable to assume that even for small $\delta p_{i}$ and $\delta \rho_{ij}$, different samples will yield significantly different topologies and $\alpha_{i}$ and $\beta_{ij}$ parameters, resulting in potentially large $\delta \alpha_{i}$ and $\delta \beta_{ij}$.

Our position on that issue is as follows: since we have by hypothesis gathered all the possible empirical information on our portfolio (summarized in probabilities of default and default correlations), and since we have argued Maxent picks the Jungle model as the credit risk model of choice in consistency with that data, and since the uncertainty on our empirical information is unavoidable, all those different models are to be considered.

As a consequence, model risk analysis would be adamant to analyse not only "the" model consistent with our empirical data, but all the models consistent with our empirical data (probably, there are many of them). This is not only a theoretical argument. For example, the Diamond model case shows that for a not too unreasonable set of probability of default and default correlation data, the probability distribution of the losses suffers a dramatic transformation (a quasi phase transition) when changing smoothly the empirical variables, especially the default correlation.

If one of the theories consistent with our empirical data were a theory having a quasi phase transition point in the vicinity of the parameters $\alpha_{i}$ and $\beta_{ij}$ consistent with our empirical data $p_{i}$ and $\rho_{ij}$, by disregarding that model we would be inadvertently creating a significant model risk.

This way of thinking is consistent with "what if" scenario analysis: it does not matter so much precision (i.e., being able to derive the "correct" $\alpha_{i}$ and $\beta_{ij}$ consistent with our empirical data $p_{i}$ and $\rho_{ij}$), but robustness, i.e., we know that our data gathering process is imperfect and we know our Jungle inverter may not be able to find always the "right" $\alpha_{i}$ and $\beta_{ij}$ consistent with our empirical data $p_{i}$ and $\rho_{ij}$; for this reason, we consider a set of potential future variations of the empirical parameters (possibly hard coded through "expert opinion") and we analyse what would happen in case those scenarios are realized.

Finally, we would like to analyse when our procedure to select the Jungle model as a relevant credit risk model, the Maxent principle under a set of empirical constraints, could fail to be a valid one. In other words, we would like to make "model risk on model risk".

The underlying hypothesis in this paper (apart from the conditions on empirical data outlined above) is we can model credit portfolios without the need to resort to the underlying, "microscopic" dynamical processes.

In particular, this hypothesis is implicit when we apply Maxent to a given credit portfolio, provided the empirical data, consisting of default probabilities and default correlations, can be obtained as follows:
 
 \begin{itemize}
 	\item $p_{i}, \forall i \in \theta$, with $p_{i} \in [0,1]$
 	
 	\item $\rho_{ij}, \forall (i,j) \in \phi$, with $\rho_{ij} \in [-1,1]$; we define $q_{ij}$ such that the relationship $\dfrac{q_{ij} - p_{i} p_{j}}{\sqrt{p_{i} (1-p_{i})} \sqrt{p_{j} (1-p_{j}) }} = \rho_{ij}, \forall (i,j) \in \phi$ holds
 \end{itemize}
 
Maxent leads to the following empirical constraints:
 
 \begin{itemize}
 	\item $p_{i} = \langle l_{i} \rangle$, $\forall i \in \theta$
 	
 	\item $q_{ij} = \langle l_{i} l_{j} \rangle, \forall (i,j) \in \phi$
 \end{itemize}

The underlying hypothesis is that during the time frame in which $p_{i}$ and $\rho_{ij}$ are fixed numbers (i.e., a time period below the typical time frames of change in those empirical variables), the "microscopic" variables $l_{i}$ fluctuate fast enough in order to be able to sample the whole space of states, and generate a meaningful value for both $\langle l_{i} \rangle$ and $\langle l_{i} l_{j} \rangle$.

If that is not the case, i.e. if the "microscopic" variables $l_{i}$ fluctuate slowly, or conversely, the empirical variables $p_{i}$ and $\rho_{ij}$ fluctuate fast (in comparison to each other), the Maxent results do not need to hold in practice.

It seems reasonable to assume that under a "good" state of the economy, the $p_{i}$ and $\rho_{ij}$ empirical values will fluctuate smoothly. Also, a credit portfolio with a low degree of default correlations (such that a binomial provides a good approximation for it) will probably "relax" fast towards its equilibrium configuration. As a consequence, a not-too-correlated portfolio, under "good" economic conditions, will probably satisfy the implicit conditions in Maxent, so the Jungle model framework will hold.

However, under a "bad" state of the economy, the $p_{i}$ and $\rho_{ij}$ empirical values will probably suffer strong and sudden fluctuations. Also, a credit portfolio with a high level of default correlations might "relax" slowly towards its equilibrium configuration. For example, it could be the dynamical processes (unknown to us) generate a state space with many local minima, and the system might be trapped in a local minimum which is not the global one, and the more the time passes, the more likely it becomes that eventually the system jumps out of that local minimum towards the another local minimum, searching the global one. In that case, the averages we would measure, $\langle l_{i} \rangle$ and $\langle l_{i} l_{j} \rangle$, would not be measurements on the whole probability distribution, but only on a small part of the state space, making the overall effort worthless (or even outright dangerous, for macroprudential purposes).

As a consequence, a highly correlated portfolio, under "bad" economic conditions, may not be correctly described by Maxent, so the Jungle model framework will not necessarily hold true.

To sum up: 

\begin{itemize}
\item The Jungle model may not be the "best" possible credit portfolio model (whatever "best" means), but at least, the Jungle model is selected by Maxent to be the credit portfolio model of choice, in consistency with the available empirical data. 
\item Model risk under the Jungle model should be thought of as an ensemble of Jungle models, defined by $\alpha_{i} \pm \delta \alpha_{i}$ and $\beta_{ij}\pm \delta \beta_{ij}$, which are consistent with the empirical data $p_{i} \pm \delta p_{i}$ and $\rho_{ij} \pm \delta \rho_{ij}$. In particular, $\delta \alpha_{i}$ and $\delta \beta_{ij}$ could be large, even for small $\delta p_{i}$ and $\delta \rho_{ij}$. An analyst should study the possibility of the existence of double peaks and quasi phase transitions for the corresponding models, since a small change in the underlying empirical data, i.e. small $\delta p_{i}$ and $\delta \rho_{ij}$, may lead to sudden and dramatic systemic events \footnote{In the same way we could not know water at 99 degrees Celsius was going to become steam when increasing temperature by one more degree.}.
\end{itemize}

\section{Modelling inhomogeneous portfolios and recovery rates}

In Section 3, we have discussed the modelling of credit losses for a general credit portfolio, including not only stochastic $l_{i}$ bi-valued indicators, but also possibly state-dependent recovery rates, and inhomogeneous exposure at default values.

In Section 4, we have simplified the above general case by only considering stochastic $l_{i}$ bi-valued indicators.

We now show the general case of Section 3 can be handled with the Jungle probability distributions of Section 4. In other words, the problem of obtaining the general probability distribution for the losses of an inhomogeneous portfolio with state-dependent recovery rates can be decoupled into two smaller problems:

\begin{itemize}
	\item First, find the Jungle probability distribution for the losses of the corresponding homogeneous portfolio with no recovery rate modelling (using the empirical probabilities of default and default correlations of the borrowers in the portfolio)
	\item Once the Jungle probability distribution is found, model the general case of the inhomogeneous portfolio with state-dependent recovery rates, by sampling the Jungle probability distribution with Markov Chain Monte Carlo methodology, and calculate trivially the corresponding losses in the general case for each realization of the Jungle probability distribution
\end{itemize}

Let us show the procedure outlined above for both inhomogeneous portfolios with no recovery rate modelling, and for homogeneous portfolios with state-dependent recovery rates, before handling the general case.

\subsection{Modelling inhomogeneous portfolios, no modelling for recovery rates}

First, let us consider the intermediate case of a credit portfolio being modelled by stochastic $l_{i}$ bi-valued indicators and inhomogeneous exposure at default values, but no recovery rates. In that case, the total loss of the portfolio can be described as:

\begin{equation}
L = \sum\limits_{i=1}^N{L_{i}} = \sum\limits_{i=1}^N{E_{i} l_{i}}
\end{equation}

The probability distribution for $L$ can be computed numerically, using the equation above, by using the probability distribution arising from the corresponding homogeneous Jungle model.

\subsection{Homogeneous portfolios with state-dependent recovery rates}

The Jungle model can also handle the case of state-dependent recovery rates for homogeneous portfolios, when the recovery rates follow the stylized fact of being lower when the overall default rate increases (and vice versa), see \citep{mora12}.

For simplicity purposes (and without loss of generality, as shown in the next subsection), let us assume a linear dependence of the recovery rate with the overall default rate, $1-RR = \frac{1 + \frac{\ell}{p}}{2}$,  where $\ell$ is $\frac{\sum\limits_{i=1}^N{l_{i}}}{N}$ and $p$ being the expected value of $\ell$.  

The expected value of $1-RR$ is one. However, since the recovery rate decreases when the default rate increases, the total loss $L$ will show the non-trivial correlations between the recovery rate and the default rate through an increase in the high loss region (with respect to the case of no state-dependent recovery rate).

\subsection{Inhomogeneous portfolios with state-dependent recovery rates}

In general, a real world portfolio will be inhomogeneous, and the recovery rates of its constituents will be state-dependent, possibly in a specific way for each borrower.

This general case is amenable to computation with the MCMC methods outlined above, and the generalization is straightforward for:

\begin{itemize}
	\item Inhomogeneous portfolios
	\item State-dependent recovery rates such as $1-RR = \frac{1 + \frac{\ell}{p}}{2}$
	\item State-dependent recovery rates with a more general functional than $1-RR = \frac{1 + \frac{\ell}{p}}{2}$, for example with non-linear terms in $l$, or with borrower specific coefficients
	\item State-dependent recovery rates that depend not only on $\ell$, but on the states of individual borrowers, $l_{i}$, for the $i$-th borrower
	\item In general, MCMC allows to compute the probability distribution of any function whose domain is the state space, $\langle f(l_{1},l_{2},\cdots,l_{N}) \rangle$
\end{itemize}

In particular, we want to highlight another suggestive possibility: $1-RR = a + b l_{0}$, with $l_{0}$ being the indicator of the central node in a Dandelion model. We have discussed above the Dandelion model introduces a relationship between macroeconomic risk factors and contagion, unifying both of them. In particular, macroeconomic risk factors could be understood as a specific, large Dandelion effect. 

As a consequence, a functional form of $1-RR = a + b l_{0}$ would mean that in the "good" state of the economy, with $l_{0} = 0$, the loss given default would be "low" ($a$, say 20\%). Instead, in the "bad" state of the economy, with $l_{0} = 1$, the loss given default would be "high" ($a+b$, say 70\%). Again, such a modelling is amenable to MCMC calculations for the corresponding Jungle model.

\section{Contagion, macroeconomic risk factors and frailty}

In this section, we point out the three factors contributing to default clustering (macroeconomic risk factors, contagion and frailty) can be understood under the unifying framework of contagion.

In particular, we show macroeconomic risk factors can be modelled as a particular case of contagion, generalizing the "Dandelion trick".

Also, we motivate frailty can be interpreted as an instance of contagion: when suddenly a hidden risk factor is revealed to the market, the effect is an abrupt and discontinuous change of the empirical parameters defining the credit portfolio, i.e. its probabilities of default and default correlations. Frailty can be thought of as the effect of this jump on the parameters of the loss probability distribution.

\subsection{Macroeconomic risk factors as contagion}

Earlier, we have seen the Dandelion model can be understood as a mixture of binomials, whereby the probability of default of a node in the periphery of the Dandelion can be decomposed as a mixture of "good" and "bad" probabilities of default, each corresponding to the two binomials in the mixture.

Here, we want to show the "Dandelion trick" can be generalized for the Jungle case, to generate a mixture not only of binomial distributions, but of any distribution arising from an interacting model: 

Given a Jungle model of $N$ credit instruments, defined by $p_{i}, i=1,\cdots,N$, and $\rho_{ij}, i=1,\cdots,N, \, j > i$, the corresponding probabilities of default and default correlations of its constituents, we can define a mixture model by:

\begin{itemize}
	\item For a fraction of time $p_{0}$, the probabilities of default of the $N$ credit instruments are $p_{1}^{BAD}, \cdots, p_{N}^{BAD}$
	\item For a fraction of time $1 - p_{0}$, the probabilities of default of the $N$ credit instruments are $p_{1}^{GOOD}, \cdots, p_{N}^{GOOD}$
\end{itemize}

By extending straightforwardly, but rather lengthily, the "Dandelion trick" for a general Jungle model, we can create mixture models of credit portfolios in a natural way. As a consequence, mixture models arising from the variation over time of macroeconomic risk factors can be embedded naturally into the framework of contagion.

\subsection{Frailty as contagion}

In this subsection, we want to motivate the use of contagion to explain, at least pictorially, the phenomenon of frailty:

Frailty is described as the "Enron effect": when "Enron" was considered to be a good company, its probability of default was low. However, some day the market discovered "Enron" had "cooked its books". Immediately, "Enron"'s probability of default sky-rocketed. 

But the impact of that discovery did not end there: analysts started wondering if other companies, in other geographic regions and in other economic sectors, probably completely unrelated to "Enron", had been cooking their books, too.

As a consequence, once the corresponding "hidden factor" was "revealed" to the market, the probabilities of default of many different companies throughout the "credit network" jumped upwards, in analogy to what happened to "Enron"'s probability of default. This behaviour is not unlike asthma (pollution may trigger a common response among people suffering asthma; but asthma is not like flu, the prototype of direct contagion).

The net result of the "Enron effect" was the probabilities of default of a multitude of nodes in the "credit network" increased suddenly. The reason was not direct contagion, but the revelation of new information to the market which had been hidden so far.

We could argue then that dealing with frailty, in the contagion framework outlined by the Jungle model, is adamant to applying "what if" scenario analysis to a given configuration of the "credit network". By definition, we cannot know what the "hidden factors" are. However, we know that whatever they are, their effect once they are discovered by the market, is a sudden jump for some of the empirical $p_{i}$ and $\rho_{ij}$. 

Frailty can then be understood as the change in the loss probability distribution of a Jungle model, when the empirical $p_{i}$ and $\rho_{ij}$ transform into a set of stressed values throughout the network, which can be quantified by the derivative of the model parameters, $\alpha$ and $\beta$, with respect to the empirical parameters, $p_{i}$ and $\rho_{ij}$.

As a consequence, frailty and contagion are intimately related, since the effect of frailty will largely depend on the contagion structure of the credit network.

\section{Policy implications of contagion}

In Section 5, we have described how the Jungle model depends on the probabilities of default and default correlations of its constituents, plus the topology of the "contagion network". 

We have seen that for several topologies, with not-too-unreasonable values for the probabilities of default and default correlations, the probability distributions of the credit losses become doubly peaked, out of credit avalanches triggered by contagion.

In particular, we have analysed how increasing the default correlation for the Dandelion model, leads to the second peak moving to more extreme losses (more extreme domino effects), as well as the first peak moving towards zero losses. The inability of some credit portfolio models to accommodate these stylized facts, even for some models used in practice for regulatory purposes such as \citep{vasicek87}, has been highlighted by \citep{kupiec09}.

In the following sections, we will describe the policy implications suggested by the effects above.

\subsection{The U.S. subprime and the European sovereign crises as quasi-phase transitions}

Both the U.S. subprime and the European sovereign crises caused sharp spikes in probabilities of default "across the board". However, such increases were mostly concentrated in specific sectors of the economy (e.g., financial corporates, such as Lehman and AIG, during the U.S. subprime crisis).

For the following argument, it will be important to highlight that not only the probabilities of default jumped during the financial crises, but also the (average pairwise) default correlations, as it can be seen from Figure 8.

\begin{figure}[ht!]
	\centering
	\includegraphics[width=95mm]{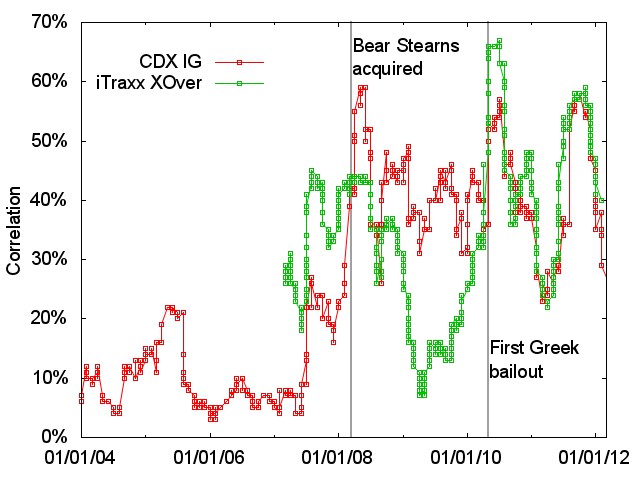}
	\caption{Historical default correlation for both iTraxx XOver and CDX IG}
\end{figure}

Before the crisis, market expectations implied low probabilities of default and default correlations for the overall economy, in particular the same held true for the financial sector. Also, risk aversion was probably "low", and as a consequence, the risk premium associated to the market values of both empirical parameters was also "low".

However, when the crisis erupted, both probabilities of default and default correlations spiked up for the financial sector as a whole. Even more, if it were possible to define the sub-sector of the most speculative parts of the financial sector (those companies exposed the most to subprime assets), it seems reasonable to assume the probability of default of such sub-sector reached exceedingly high levels, of the order of magnitude of $50\%$.

From the analysis of the Diamond model, we have seen quasi phase transitions arise naturally. And the Diamond model, with equally pairwise default correlations, is the corresponding Jungle model associated to a homogeneous portfolio whereby all nodes are connected to each other. And the sub-sector of the most speculative parts of the financial sector constitutes a credit portfolio which probably can be correctly approximated as homogeneous, with each node connected to any other node in the network. In fact, the standard methodology to imply default correlations from traded credit indices, such as CDX IG or iTraxx XOver, assumes all their individual components have the same default correlation with every other name in the index.

Then, for a Diamond model, and under "normal" (non-crisis) conditions, the fundamental situation of the economy is such that both the probabilities of default and the default correlations are "low", resulting in an economy located around the bottom-left corner of Figure 7. However, in a crisis, and for that sub-sector of the economy exposed the most to the key fundamentals of the crisis, the model parameters $\alpha$ and $\beta$ may get closer to the "line of quasi phase transitions", from below. As we have observed before, the "critical point" of such line corresponds to a probability of default of $44\%$ and a default correlation of $11\%$ for a reasonable set of parameters.

As a consequence, if for such a sub-sector, the impact of the deterioration in the fundamental situation of the economy implies that the corresponding probabilities of default and default correlations are close (from below) to $44\%$ and $11\%$, resp., just a little bit of further economic deterioration may result in the model parameters, $\alpha$ and $\beta$, crossing up the "line of quasi phase transitions" (from the bottom left corner, to the top right corner). Such a small change may be immaterial in the model parameters space, but in the empirical parameters space, the impact is huge: both the probability of default and the default correlation spike up, from low levels (close to $0\%$) to high levels (close to $100\%$).

The policy implication of the discussion above is then as follows: monitor closely the model parameters, $\alpha$ and $\beta$, of the sectors of the economy most exposed to potential future systemic crises (the financial sector being always one of such sectors), and raise a red flag once and if the model parameters are getting closer to the "line of quasi phase transitions", or at least, a reasonable stress test suggests the model parameters might cross such line in case of a sudden macroeconomic / microeconomic shock.

In particular, let us highlight that such a policy would have resulted in a "red flag" for both the U.S. subprime and the European sovereign crises at the beginning of such crises, and even possibly a bit before their sudden eruption, as it can be seen intuitively from the default correlation figures above.

\subsection{Understanding the historical probability distributions of credit losses}

The historical default rates provided by \citep{moodys} (also data from \citep{giesecke11} would allow us to reach similar conclusions), presented in Section 2, yield the following histograms:


\begin{figure*}[!ht]
	\centering
	\includegraphics[width=4.9cm]{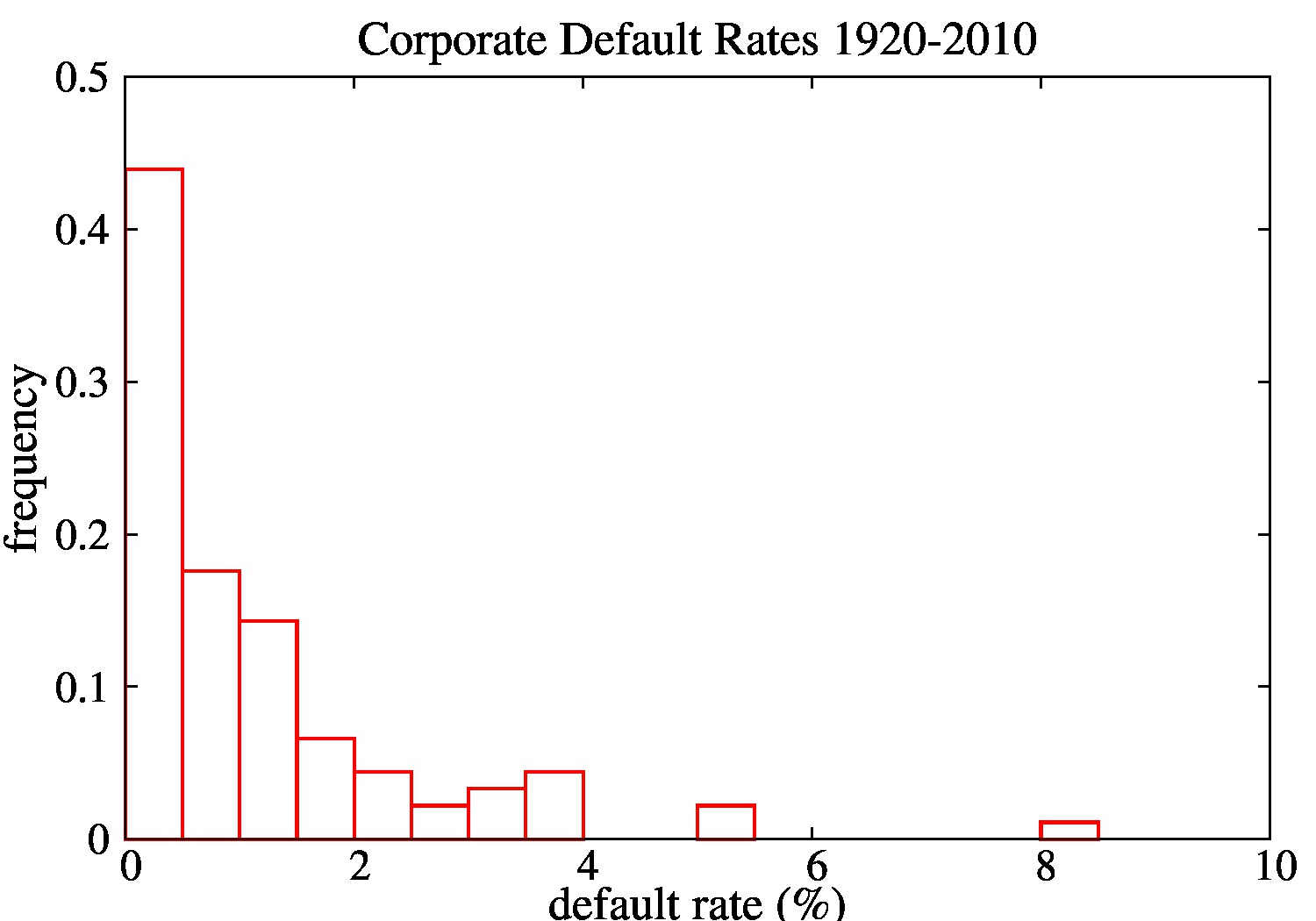}%
	\qquad
	\includegraphics[width=4.9cm]{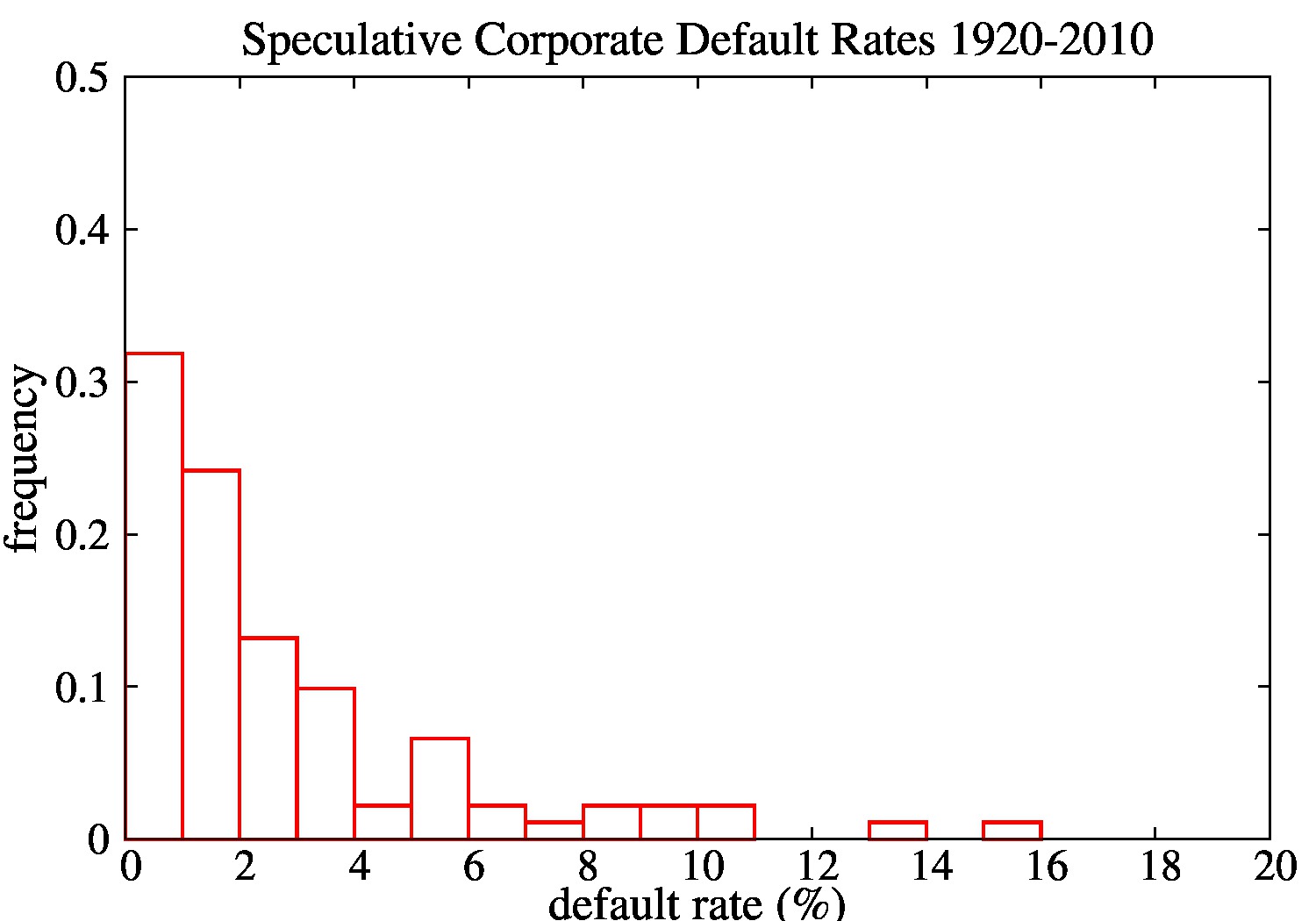}%
	\qquad
	\includegraphics[width=4.9cm]{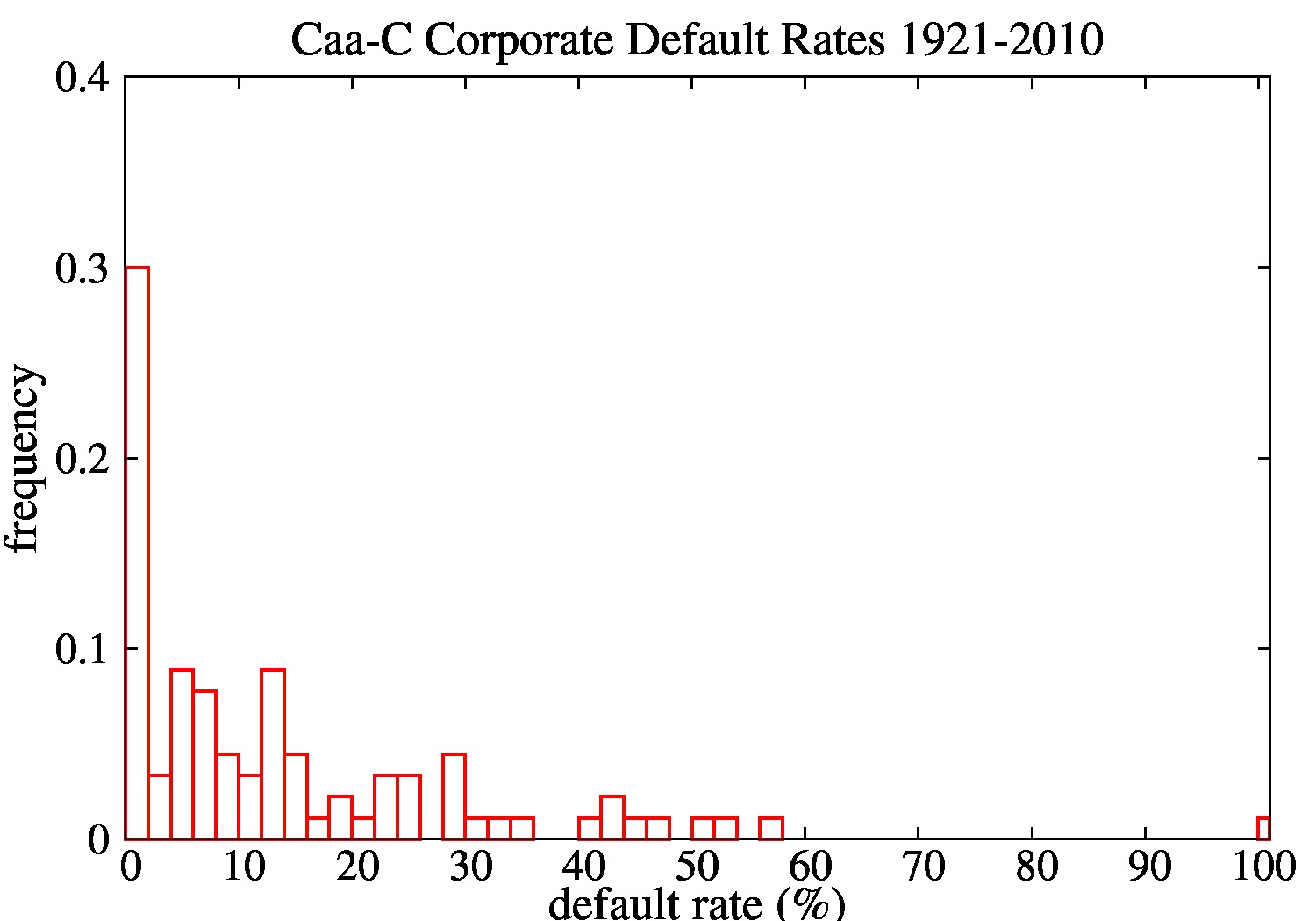}%
	\qquad
	
	\caption{Histograms for data in \citep{moodys}}
\end{figure*}

From visual inspection for the three figures above, there are too few data points to robustly ascertain if the probability distributions for the default rates have one peak or more. Intuitively, it seems the tail is a fat one, with credit loss realizations up to 100\%, in the case of Caa-C rating.

The question is then if this evidence contradicts our claim that the fact that Maxent picks the Jungle model as its probability distribution of losses of choice suggests the Jungle model is a reasonable credit risk model to be considered in practice, since the Jungle model is often (for several topologies) doubly peaked, as for example the Dandelion model.

The answer is that it does not. The historical distribution of losses presented above can be understood as follows:

Let us assume, without loss of generality, that the empirical probabilities of default and default correlations only change once per year, on Jan 1st. Let us assume the corresponding topology gives rise to a Jungle model generating doubly-peaked probability distributions. Then, the losses that year will be a realization of that particular Jungle model. Probably, the realization will fall under the first peak. But the more years we repeat the same procedure (with their corresponding probabilities of default, default correlations and topology), the more likely is a realization occurs on the second peak (contagion effects, generating an avalanche / domino effect of credit defaults).

From the Dandelion model, we have found out the position of the second peak is largely determined by the default correlation (the probabilities of default also matter).

As time passes by, we will have a series of realizations of the second peak. But importantly, the empirical data for each realization (probabilities of default, default correlations and topology) will most likely be different for each year, probably generating a double peak at different location on the axis of losses for each realization of the second peak.

As a consequence, the historical probability distribution of losses will probably have only a first peak, consistent with the fact that in the majority of realizations, losses are basically contagion-less, so that first peak will be roughly similar to the one of a binomial model, but wider due to the mixing with different macroeconomic conditions over several business cycles, and a fat tail generated by realizations of the doubly-peaked probability distributions arising from the Jungle model.

This way of thinking allows us to understand how is it possible the tail of the empirical probability distributions is so "thick": the tail is generated through individual realizations of double peaks. This way of thinking relaxes the need to include extreme probability distributions which are able to cope by themselves with the difficult task to model both extreme default events, and default events in a "good" economy state.

Even more, the Jungle model allows us to understand a stylized fact of the probability distribution of losses for highly risky portfolios, exemplified by Moody's Caa-C rating data: despite the fact Caa-C rating bonds are highly risky (and there is even one year where 100\% of bonds in the sample defaulted), it also happens very often that Caa-C rating bonds enjoy a default rate close to 0\% (on the sample, there are several years with 0\% default rate). In fact, this phenomenon of 0\% default rate happens more often for Caa-C than for bonds with a much better rating, which seems intuitively odd, see \citep{kupiec09}.

However, the Dandelion model is able to explain this stylized fact: for Caa-C rating bonds, it seems likely that the individual bonds are described not only by high probabilities of default, but also by high default correlations among themselves (or with a central node, in a similar way to the Dandelion model; possibly banks or other financial suppliers specialized on risky lending). 

From the charts in the Dandelion model section, we can see that in this region of parameters, the higher the default correlation, the larger the losses for the double peak. But in addition to this effect, also the higher the default correlation, the lower the losses for the first peak. This is consistent with a contagion effect: contagion not only works on "bad" situations (a default in a node induces a default in another node nearby), but also on "good" situations (a non-default in a node induces a non-default in another node nearby).

As a consequence, this framework of thinking leads us to suggest that the most relevant variable to ascertain default clustering is not the probability of default (as standard rating classifications appear implicitly to suggest) but default correlations.

\subsection{How should the Jungle model be used in practice?}

The Jungle model could be extended straightforwardly to introduce macroeconomic risk factors. The Jungle model is based under the assumption that the probabilities of default and default correlations are known fixed numbers. An analyst could include into the modelling a specific pattern of conditional default probabilities (and possibly conditional default correlations as well), in the same way mixture models introduce default correlations by mixing the binomial model (independent defaults) with a specific choice of conditional default probabilities.

However, the interest to do so is limited. On one hand, the Jungle model does not need to follow the above procedure to include default correlations, since default correlations appear endogenously in the model. This is unlike the case for the binomial model, which needs a mixture in order to be able to model default clustering. On the other hand, averaging over a conditional default probability means the corresponding probability distribution of losses is the one for the whole business cycle. 

We believe a probability distribution of credit losses for a full business cycle is of limited interest, since a macroprudential regulator or a bank senior risk manager is interested mainly in extreme events. So, what matters for the probability distribution of losses is the short future, if and when a "bad" economic scenario is realized.

As a consequence, the Jungle model fits better with a "what if" strategy for model risk: "What would happen to my credit portfolio if my topology, probabilities of default and default correlations change?" The reason is the Jungle model is the most general probability distribution for credit losses (under the assumptions presented in the Introduction and later sections), so changing the parameters allow the analyst to span the whole space of credit models.

By exploring how the probability distribution of credit losses for the Jungle model changes once the topology, the probabilities of default and the default correlations change (either by econometric equations, or by expert knowledge), an analyst could get a robust feel for credit tail risks.

\subsection{It's the correlations, stupid!}

Using a pictorial, non-rigorous example, the above discussion suggests an economy full of "industrial entrepreneurs", understood as risky projects on a standalone basis, but lowly correlated among each other, will probably have a lower systemic risk than an economy full of "financial large corporates", understood as corporates with low probabilities of default, but highly correlated among each other (possibly through a strong, common dependence on financial conditions).

A credit portfolio of "industrial entrepreneurs" will probably have a high expected loss and a wide first peak (which would intuitively suggest a high risk portfolio), but probably will have a small or negligible second peak, since domino effects will probably not be present.

Instead, a credit portfolio of "financial large corporates" will probably have a low expected loss and a narrow first peak (which would intuitively suggest a low risk portfolio), but probably will have a significant second peak, since domino effects might be relevant if the common, "financial" factor underlying the performance of all those large corporates suffers a severe, unexpected, "Black Swan" crisis, \textit{\`a la} \citep{taleb07}.

As a consequence, measuring credit risk in a bottom-up fashion, for example assuming a given credit portfolio is "low risk" because its individual constituents are "low risk" (in the sense of having a good credit rating), disregarding collective effects at the portfolio level, may severely understate the systemic risks of a credit portfolio as a whole.

\subsection{"Too Big To Fail" banks}

Similarly, the discussion above suggests the "Too Big To Fail" (TBTF) phenomenon for large banks, a relevant issue for macroprudential regulators, can be understood within our contagion framework. In particular, a portfolio consisting of a TBTF bank and its borrowers can often be modelled as a Dandelion model. 

The Dandelion model results show the probability distribution of losses for the Dandelion suffers from a significant double peak even for non-extreme probabilities of default and default correlations' estimates. This double peak may lead to contagion effects when the TBTF bank is close to bankruptcy, so the regulator may have the incentive to "bail out" the TBTF bank when the regulator realizes the total losses out of the TBTF bank default (implying massive defaults on its borrowers, due to domino effects out of contagion) are simply not socially acceptable. A TBTF bank knows the regulator knows that, so the incentive for the TBTF bank is to grow as much as it can, and to create as much contagion as it can. This is an example of social and economic fragility, as described in \citep{taleb12}.

A rational regulator should be able to understand these perverse incentives, and pre-emptively regulate banks' size, with a penalty for large sizes.

As described above, the onus of the argument for bank regulation is not to be found by analysing the TBTF bank on an individual basis, but understanding the effect is an emerging, global one. In other words, the "capitalist freedom" of a bank to grow as much as it can is interfering with the "capitalist freedom" of other economic agents in the economy without a say on the banks' actions, but potentially suffering the severe consequences of a systemic crises when and if the TBTF defaults and creates an avalanche of defaults through contagion (capitalism should be an incentive system where one is rewarded or punished by the result of one's actions, not by the actions of somebody else). 

This phenomenon is clearly an externality, and it may require proper regulation to safeguard the interests of society as a whole.

\section{Conclusions}

This paper presents and develops the Jungle model in a credit portfolio framework. The Jungle model generates a probability distribution for the losses of a credit portfolio with the following stylized facts:

\begin{enumerate}
	\item The Jungle model is able to model contagion among borrowers
	\item The Jungle model endogenously generates doubly-peaked probability distributions for the credit losses, with the second peak consistent with avalanches / domino effects out of contagion
	\item The Jungle model endogenously generates quasi phase transitions, meaning small changes in the portfolio may result in sudden and unexpected systemic risks. The Jungle model helps us to ascertain the location and nature of those quasi phase transition points
\end{enumerate}

We study a series of particular cases of the Jungle model, in particular the Dandelion model and the Diamond model.

The Dandelion model exemplifies the emergence of doubly-peaked probability distributions. The Diamond model quantifies how and when a quasi phase transition may occur for the Jungle model.

Model risk arises from the genuine model uncertainty: potentially, there will be many different Jungle models consistent with the set of available empirical data for our portfolio. As a consequence, by considering the potential systemic risks of this ensemble of Jungle theories allows us to address model risk. In particular, we have shown that for not too unreasonable data, some Jungle models endogenously generate a quasi phase transition, i.e. given small changes in the underlying empirical parameters may induce sudden changes in the collective behaviour of the system, potentially and inadvertently generating systemic events. Quasi phase transitions and doubly-peaked probability distributions represent a challenge for model risk.

We show the Jungle model is able to handle inhomogeneous portfolios and state-dependent recovery rates.

We argue the Jungle model provides a unifying framework to think about the three factors contributing to default clustering (macroeconomic risk factors, contagion and frailty).

The analysis of the Jungle model in general, and of the Dandelion and Diamond models in particular, leads to some policy implications of contagion. We are able to understand qualitatively some empirical evidence, such as the embedding of the U.S. subprime and the European peripheral crises into the general framework of quasi phase transitions, the thick tails in the historical probability distributions, as well as the surprising fact that quite often, the worst quality credit portfolios end up with default rates lower than the corresponding ones with a better rating. We also analyse the "Too Big To Fail" phenomenon under our framework based on contagion, and we pictorially compare systemic risks out of contagion for a "financial economy of big corporates" versus an economy of "industrial entrepreneurs".

We believe the study of the Jungle model in the credit arena, especially for regulatory purposes, deserves further attention.

\bigskip
\noindent
	
\appendices
\section{The Maxent principle}

Maxent asserts:

Given a finite state space $\Omega$, the probability distribution in $\Omega$ that maximizes the entropy and satisfies the following $m < \mathbf{card}(\Omega)$ constraints, given $m$ different functions in $\Omega$, $f_{k}(x)$, and $m$ fixed numbers $F_{k}$:

\begin{equation}
\langle f_{k}(x) \rangle := \sum\limits_{x \in \Omega} P(x) f_{k}(x) = F_{k}, \quad k=1,2,\cdots,m
\end{equation}

as well as a normalization condition:

\begin{equation}
\langle 1 \rangle := \sum\limits_{x \in \Omega} P(x) = 1
\end{equation}

is:

\begin{equation}
P(x) = \dfrac{1}{Z(\lambda_{1}, \lambda_{2}, \cdots, \lambda_{m})} \exp{\left(- \sum\limits_{i = 1}^m \lambda_{i} f_{i}(x)\right)}
\end{equation}

where $Z$ is called the partition function:

\begin{equation}
Z(\lambda_{1}, \lambda_{2}, \cdots, \lambda_{m}) = \sum\limits_{\Omega} \exp{\left(- \sum\limits_{i = 1}^m \lambda_{i} f_{i}(x)\right)}
\end{equation}

The Lagrange multipliers $\lambda_{i}$ are found by inverting the following set of $m$ equations:

\begin{equation}
F_{k} = \langle f_{k}(x) \rangle = - \frac{\partial \, \log Z(\lambda_{1}, \lambda_{2}, \cdots, \lambda_{m})}{\partial \, \lambda_{k}} , \quad k=1,2,\cdots,m
\end{equation}

Let us prove the claim above:

Entropy is a function on $\Omega$ defined by:

\begin{equation}
S(p_{1}, \cdots, p_{n}) := - \sum\limits_{i=1}^n p_{i} \log(p_{i})
\end{equation}

where we define $n := \mathbf{card}(\Omega)$.

As a consequence, we want to find the probability distribution $p_{i}, \quad i = 1, \cdots, n$ such that $S$ is maximum for that choice of $p_{i}$ over any other possible choice of $p_{i}$, under the constraints:

\begin{itemize}
	\item $\sum\limits_{i=1}^N p_{i} = 1$
	\item $\langle f_{k}(x) \rangle = \sum\limits_{i=1}^n p_{i} f_{k}(x_{i})= F_{k}, \quad k=1,2,\cdots,m$ and $x_{i}$ denotes the $i$-th state
\end{itemize}

First, we can find an extremum of $S$ through the method of Lagrange multipliers, by imposing:

\begin{align}
0 = \delta \left[S - (\lambda_{0} - 1) \sum\limits_{i=1}^n p_{i} - \sum\limits_{k=1}^m \lambda_{k} \sum\limits_{i=1}^n p_{i} f_{k}(x_{i})\right] \\
= \sum\limits_{i=1}^n \left[ \dfrac{\partial S}{\partial p_{i}} - (\lambda_{0} - 1) - \sum\limits_{k=1}^m \lambda_{k} f_{k}(x_{i}) \right] \delta p_{i}
\end{align}

After some algebra, we find:

\begin{equation}
p_{i} = \exp \left(-\lambda_{0} -  \sum\limits_{k=1}^m \lambda_{k} f_{k}(x_{i}) \right)
\end{equation}

By imposing the constraint that all probabilities must sum up to one, we obtain:

\begin{equation}
\exp \left(\lambda_{0}\right) = \sum\limits_{i=1}^n \exp\left(-\sum\limits_{k=1}^m \lambda_{k} f_{k}(x_{i})\right)
\end{equation}

By defining $Z := \exp \left(\lambda_{0}\right)$, we obtain:

\begin{equation}
P(x) = \dfrac{1}{Z(\lambda_{1}, \lambda_{2}, \cdots, \lambda_{m})} \exp{\left(- \sum\limits_{k = 1}^m \lambda_{k} f_{k}(x)\right)}
\end{equation}

And the other constraints are satisfied as follows:

\begin{equation}
F_{k} = \langle f_{k}(x) \rangle = \dfrac{1}{Z(\lambda_{1}, \lambda_{2}, \cdots, \lambda_{m})} \sum\limits_{i = 1}^n f_{k}(x_{i}) \exp{\left(- \sum\limits_{j = 1}^m \lambda_{j} f_{j}(x_{i})\right)} = - \frac{\partial \, \log Z(\lambda_{1}, \lambda_{2}, \cdots, \lambda_{m})}{\partial \, \lambda_{k}}
\end{equation}

Now we only need to prove the above extremum is in fact a maximum.

Let us assume we have a set of non-negative numbers $p_{i}, \quad i = 1, \cdots, n$ such that $\sum\limits_{i=1}^N p_{i} = 1$. Let us assume the numbers $u_{i}$ satisfy the same conditions.

It is easy to derive that $\log(x) \leq x - 1$, for non-negative $x$. The equality only holds for $x = 1$. Then

\begin{equation}
\sum\limits_{i=1}^n p_{i} \log\left(\frac{u_{i}}{p_{i}}\right) \leq \sum\limits_{i=1}^n p_{i} \left(\frac{u_{i}}{p_{i}} - 1\right) = 0
\end{equation}

which is equivalent to

\begin{equation}
S(p_{1}, \cdots, p_{n}) \leq \sum\limits_{i=1}^n p_{i} \log\left(\frac{1}{u_{i}}\right)
\end{equation}

with the equality only holding for $u_{i} = p_{i}, \quad i = 1, \cdots, n$.

Now, defining:

\begin{equation}
u_{i} := \dfrac{1}{Z(\lambda_{1}, \lambda_{2}, \cdots, \lambda_{m})}  \sum\limits_{k=1}^m \exp\left(-\lambda_{k} f_{k}(x_{i})\right)
\end{equation}

The inequality above for the entropy can be written as:

\begin{equation}
S(p_{1}, \cdots, p_{n}) \leq \sum\limits_{i=1}^n p_{i} \left[ \log Z(\lambda_{1}, \lambda_{2}, \cdots, \lambda_{m}) + \sum\limits_{k=1}^m \lambda_{k} f_{k}(x_{i}) \right] = \log Z(\lambda_{1}, \lambda_{2}, \cdots, \lambda_{m}) + \sum\limits_{k=1}^m \lambda_{k} \langle f_{k}(x_{i}) \rangle
\end{equation}

Now, let us consider all possible probability distributions $p_{i}, \quad i = 1, \cdots, n$ that satisfy the initial constraints. In the equation above, the right hand side is a constant for all these probability distributions. Instead, the left hand side (the entropy) changes when changing the probability distribution, and its maximum only is attained (i.e., the inequality becomes an equality) when $p_{i}, \quad i = 1, \cdots, n$ becomes the following distribution:

\begin{equation}
p_{i} = \dfrac{1}{Z(\lambda_{1}, \lambda_{2}, \cdots, \lambda_{m})}  \exp \left(-  \sum\limits_{k=1}^m \lambda_{k} f_{k}(x_{i}) \right)
\end{equation}

\section{The binomial model}

The homogeneous Jungle model with external field but no $\beta_{ij}$:

\begin{equation}
P(l_{1},l_{2},\cdots,l_{N}) = \dfrac{1}{Z} \exp{\left(\alpha \sum\limits_{i=1}^N{ l_{i}}\right)}
\end{equation}

has a partition function which can be calculated exactly:

\begin{align}
Z = \sum_{l_{1},l_{2},\cdots,l_{N}} \exp{\left(\alpha \sum\limits_{i=1}^N{ l_{i}}\right)} = \sum_{l_{1}=0}^1 \sum_{l_{2}=0}^1 \cdots \sum_{l_{N}=0}^1 \exp{\left(\alpha \sum\limits_{i=1}^N{ l_{i}}\right)} = \sum_{l_{1}=0}^1 \exp{\left(\alpha l_{1}\right)} \sum_{l_{2}=0}^1 \exp{\left(\alpha l_{2}\right)} \cdots \sum_{l_{N}=0}^1 \exp{\left(\alpha l_{N}\right)} \\
\noindent = \left(\sum_{l=0}^1 \exp{\left(\alpha l\right)}\right)^N = \left(1+e^\alpha\right)^N
\end{align}

The explicit knowledge of the partition function allows us to compute any value for the theory. For example, we can compute the expected value of the loss, and see how the $\alpha$ parameter directly relates to the empirical $p$:

\begin{equation}
p = \dfrac{\langle \ell \rangle}{N} = \dfrac{1}{N} \dfrac{\partial \log Z}{\partial \alpha} = \dfrac{\partial \log (1 + e^\alpha)}{\partial \alpha} = \dfrac{1}{1 + e^{- \alpha}}
\end{equation}

or equivalently:

\begin{equation}
e^{- \alpha} = \dfrac{1-p}{p}
\end{equation}

Now we can compute the probability distribution of the losses, recalling a probability distribution can be computed by counting in how many ways the states can be arranged to produce the given underlying variable, divided by the partition function:

\begin{equation}
P\left(\sum\limits_{i=1}^N{ l_{i}} = \ell \right) = \dfrac{1}{Z} \exp{\left(\alpha \sum\limits_{l_{1},l_{2},\cdots,l_{N} | \sum\limits_{i=1}^N{ l_{i}} = \ell}^N{ l_{i}}\right)}
\end{equation}

We recall there are $\binom{N}{\ell}$ ways to choose $\ell$ elements out of $N$. Then:

\begin{equation}
P\left(\sum\limits_{i=1}^N{ l_{i}} = \ell \right) = \dfrac{1}{Z} \binom{N}{\ell} e^{\alpha}
\end{equation}

By substituting $Z$ with its value above, and $\alpha$ by its dependence on $p$ above, we find:

\begin{equation}
P\left(\sum\limits_{i=1}^N{ l_{i}} = \ell \right) = \binom{N}{\ell} p^\ell (1-p)^{N-\ell}
\end{equation}

\section{Small contagion}

We have seen in Appendix B that the Jungle model:

\begin{equation}
P(l_{1},l_{2},\cdots,l_{N}) = \dfrac{1}{Z} \exp{\left(\alpha \sum\limits_{i=1}^N{ l_{i}} + \sum\limits_{(i,j) \in \phi}{\beta_{ij} l_{i} l_{j}}\right)}
\end{equation}

becomes a binomial distribution when $\beta_{ij} = 0$, $\forall i,j$. In other words, the Jungle model with no $\beta$:

\begin{equation}
P_{\beta=0}(l_{1},l_{2},\cdots,l_{N}) = \dfrac{1}{Z} \exp{\left(\alpha \sum\limits_{i=1}^N{ l_{i}}\right)}
\end{equation}

becomes the binomial distribution:

\begin{equation}
P\left(\sum\limits_{i=1}^N{ l_{i}} = \ell \right) = \binom{N}{\ell} p^\ell (1-p)^{N-\ell}
\end{equation}

with $p = \dfrac{1}{1 + e^{- \alpha}}$.

Now, we are interested in the Jungle model with the same $\alpha$ as above, but also with $\beta_{12} =: \beta$ and $\beta_{ij} = 0,$ for $ij \neq 12$. In other words, we are interested in expanding perturbatively around the binomial model, in order to understand the effect of $\beta$. The corresponding Jungle model is then:

\begin{equation}
P_{\beta}(l_{1},l_{2},\cdots,l_{N}) = \dfrac{1}{Z} \exp{\left(\alpha \sum\limits_{i=1}^N{ l_{i}} + \beta l_{1} l_{2}\right)}
\end{equation}

The corresponding partition function can be calculated exactly:

\begin{align}
Z_{\beta} = \sum_{l_{1}=0}^1 \sum_{l_{2}=0}^1 \cdots \sum_{l_{N}=0}^1 \exp{\left(\alpha \sum\limits_{i=1}^N{ l_{i}} + \beta l_{1} l_{2}\right)} =\sum_{l_{1}=0}^1 \sum_{l_{2}=0}^1 \exp{\left(\alpha (l_{1} + l_{2}) + \beta l_{1} l_{2} \right)}  \left(\sum_{l=0}^1 e^{\alpha l}\right)^{N-2} \\
=\left(e^{\beta} e^{2 \alpha} + 2 e^{\alpha} + 1\right) \left(1 + e^{\alpha}\right)^{N-2}
\end{align}

Rearranging terms:

\begin{equation}
Z_{\beta} =(e^{\beta} - 1) e^{2 \alpha} (1 + e^{\alpha})^{N-2} + (1 + e^{\alpha})^{N} = \dfrac{(e^{\beta} - 1) e^{2 \alpha}}{(1 + e^{\alpha})^{2}} (1 + e^{\alpha})^{N} + (1 + e^{\alpha})^{N} = Z_{\beta=0} \left(1 + \dfrac{e^{\beta} - 1}{(1 + e^{-\alpha})^{2}}\right)
\end{equation}

The explicit knowledge of the partition function allows us to compute any value for the theory. For example, we can compute the expected value of the loss of $l_{1}$:

\begin{align}
p_{1}^{\beta} = \langle l_{1} \rangle_{\beta} = \frac{1}{Z} \sum_{l_{1}=0}^1 \sum_{l_{2}=0}^1 \cdots \sum_{l_{N}=0}^1 l_{1} \exp{\left(\alpha \sum\limits_{i=1}^N{ l_{i}} + \beta l_{1} l_{2}\right)} \\ 
= \frac{1}{Z} \sum_{l_{1}=0}^1 \sum_{l_{2}=0}^1 l_{1} \exp{\left(\alpha (l_{1} + l_{2})  + \beta l_{1} l_{2}\right)} \left(\sum_{l=0}^1 e^{\alpha l}\right)^{N-2} = \frac{1}{Z} \left(e^{\alpha} + e^{2 \alpha + \beta}\right) \left(1 + e^{\alpha}\right)^{N-2}
\end{align}

Substituting the partition function and rearranging:

\begin{equation}
p_{1}^{\beta} = \langle l_{1} \rangle_{\beta} = \frac{e^{\alpha} + e^{2 \alpha + \beta}}{e^{\beta} e^{2 \alpha} + 2 e^{\alpha} + 1} = \frac{(e^{\beta} - 1) + (1 + e^{-\alpha})}{(e^{\beta} - 1) + (1 + e^{-\alpha})^2}
\end{equation}

Taking $\beta$ small (keeping only up to linear terms in $\beta$) and substituting $1 + e^{-\alpha}$ by $\frac{1}{p_{\beta=0}}$: 

\begin{equation}
p_{1}^{\beta} = \langle l_{1} \rangle_{\beta} \approx \frac{\beta + \frac{1}{p_{\beta=0}}}{\beta + \frac{1}{p_{\beta=0}^2}} = p_{\beta=0} \frac{1 + \beta p_{\beta=0}}{1 + \beta p_{\beta=0}^2} \approx p_{\beta=0} (1 + \beta p_{\beta=0})(1 - \beta p_{\beta=0}^2) \approx p_{\beta=0} (1 + \beta p_{\beta=0} (1 - p_{\beta=0}))
\end{equation}

As a consequence, $p_{\beta}$ is slightly larger than $p_{\beta=0}$, despite both sharing the same $\alpha$, and the increase is proportional to $\beta$.

For $\beta$ not necessarily small, it still holds true that $p_{\beta}$ is larger than $p_{\beta=0}$, the only difference with the small $\beta$ case is that the increase in probability of default is not proportional to $\beta$ any more.

We can then state that small contagion effects around the binomial model are akin to increasing the probability of default of the system.

We also can see that, in addition to increase the probability of default, $\beta$ induces correlations:

\begin{equation}
\langle l_{1} l_{2} \rangle_{\beta} = \dfrac{\partial \, \log Z}{\partial \beta} = \dfrac{\partial \, \log Z_{\beta=0}}{\partial \beta} + \dfrac{\partial \, \log \left(1 +  \dfrac{e^{\beta} - 1}{(1+e^{- \alpha})^2}\right)}{\partial \beta} =  0 + \frac{1}{1+\dfrac{e^{\beta} - 1}{(1+e^{- \alpha})^2}}  \dfrac{e^{\beta}}{(1+e^{- \alpha})^2}
\end{equation}

Rearranging terms and using the dependence of $p_{\beta=0}$ with $\alpha$:

\begin{equation}
\langle l_{1} l_{2} \rangle_{\beta} =  \frac{e^{\beta}}{\frac{1}{p_{\beta=0}^2} +e^{\beta} - 1} = p_{\beta=0}^2 \frac{e^{\beta}}{1+ p_{\beta=0}^2 (e^{\beta} - 1)}
\end{equation}

Let us finally check the correlation between $1$ and $2$ is proportional to $\beta$. From:

\begin{equation}
\rho_{12}^{\beta} = \dfrac{\langle l_{1} l_{2} \rangle_{\beta} - \langle l_{1} \rangle_{\beta} \langle l_{2} \rangle_{\beta}}{\sqrt{\langle l_{1} \rangle_{\beta} (1-\langle l_{1} \rangle_{\beta})} \sqrt{\langle l_{2} \rangle_{\beta} (1-\langle l_{2} \rangle_{\beta})}}
\end{equation}
	
We only need to take care of the numerator to accomplish our goal. Let us take up to linear terms in $\beta$ for $\langle l_{1} l_{2} \rangle_{\beta}$:

\begin{equation}
\langle l_{1} l_{2} \rangle_{\beta} \approx p_{\beta=0}^2 \frac{1 + \beta}{1 - p_{\beta=0}^2 \beta} \approx p_{\beta=0}^2 (1 + \beta (1 - p_{\beta=0}^2))
\end{equation}
	
Then, up to linear terms in $\beta$:

\begin{align}
\langle l_{1} l_{2} \rangle_{\beta} - \langle l_{1} \rangle_{\beta} \langle l_{2} \rangle_{\beta} \approx p_{\beta=0}^2 (1 + \beta (1 - p_{\beta=0}^2)) - p_{\beta=0}^2 (1 + 2 \beta p_{\beta=0} (1 - p_{\beta=0})) \\
= \beta (p_{\beta=0}^2 (1 - p_{\beta=0}^2) - 2 p_{\beta=0}^3 (1 - p_{\beta=0})) = \beta p_{\beta=0}^2 (1 - p_{\beta=0})^2
\end{align}

As a consequence, the correlation is proportional to $\beta$ with a positive coefficient. 

We have finally proven that for small contagion, the coefficient $\beta$ can be interpreted as (a simple function of) the default correlation, in the same way that for no contagion, $\alpha$ can be interpreted as (a simple function of) the probability of default.

\section{The Dandelion model}

The Dandelion model corresponds to a Jungle model with $N+1$ borrowers, such that the first one, defined as $i=0$ and considered to be at the centre of the Dandelion, is "connected" to all remaining borrowers, at the external surface of the Dandelion, such that $\beta_{0i} =: \beta \neq 0$ for $i = 1, 2, \cdots, N$. Any other borrowers remain unconnected, $\beta_{ij} = 0$ for $i = 1, 2, \cdots, N \quad \& \quad j > i$. For simplicity, we assume $\alpha_{i} =: \alpha$ for $i = 1, 2, \cdots, N$.

The probability distribution for the Dandelion model is:

\begin{equation}
P(l_{1},l_{2},\ldots,l_{N}) = \dfrac{1}{Z} \exp{ \left(\alpha_{0} l_{0} + \alpha \sum\limits_{i = 1}^N l_{i} + \beta \sum\limits_{i=1}^N l_{0} l_{i}\right)}
\end{equation}

The Dandelion model, despite being interacting, can be fully solved:

\begin{align}
Z =  \sum\limits_{l_{0} = 0}^1 \sum\limits_{l_{1} = 0}^1 \cdots \sum\limits_{l_{N} = 0}^1 \exp{ \left(\alpha_{0} l_{0} + \alpha \sum\limits_{i = 1}^N l_{i} + \beta \sum\limits_{i=1}^N l_{0} l_{i}\right)}\\
 = \sum\limits_{l_{1} = 0}^1 \cdots \sum\limits_{l_{N} = 0}^1 \exp{ \left(\alpha \sum\limits_{i = 1}^N l_{i}\right) } \sum\limits_{l_{0} = 0}^1  \exp{ \left(l_{0} \left(\alpha_{0} +  \beta \sum\limits_{i=1}^N l_{i}\right)\right) }
\end{align}

Summing over $l_{0}$:

\begin{align}
Z = \sum\limits_{l_{1} = 0}^1 \cdots \sum\limits_{l_{N} = 0}^1 \exp{ \left(\alpha \sum\limits_{i = 1}^N l_{i}\right) } \left(1 +\exp{ (\alpha_{0} +  \beta \sum\limits_{i=1}^N l_{i})}\right)\\
= \sum\limits_{l_{1} = 0}^1 \cdots \sum\limits_{l_{N} = 0}^1 \exp{ \left(\alpha \sum\limits_{i = 1}^N l_{i}\right) } +e^{\alpha_{0}}  \sum\limits_{l_{1} = 0}^1 \cdots \sum\limits_{l_{N} = 0}^1 \exp{ \left((\alpha + \beta) \sum\limits_{i=1}^N l_{i}\right)}
\end{align}

In analogy with the binomial case discussed in Appendix B, the sums above can be calculated explicitly, giving:

\begin{equation}
Z = (1 + e^{\alpha})^N + e^{\alpha_{0}}  (1 + e^{\alpha + \beta})^N
\end{equation}

And then:

\begin{equation}
\log Z = N \log \left(1 + e^{\alpha}\right) + \log \left(1 + e^{\alpha_{0}}  \left(\frac{1 + e^{\alpha + \beta}}{1+e^{\alpha}}\right)^N\right)
\end{equation}

With an explicit knowledge of $log Z$, as a function of $\alpha$, $\alpha_{0}$ and $\beta$, we can derive $p$, $p_{0}$ and $\rho$ as a function of $\alpha$, $\alpha_{0}$ and $\beta$, as described in the Maxent description:

\begin{align*}
p_{0} = \langle l_{0} \rangle = \frac{\partial \, \log Z(\alpha_{0}, \alpha, \beta)}{\partial \, \alpha_{0}}\\
p = \frac{\sum\limits_{i = 1}^N \langle l_{i} \rangle}{N} = \frac{1}{N} \frac{\partial \, \log Z(\alpha_{0}, \alpha, \beta)}{\partial \, \alpha}\\
q = \frac{\sum\limits_{i = 1}^N \langle l_{0} l_{i} \rangle}{N} = \frac{1}{N} \frac{\partial \, \log Z(\alpha_{0}, \alpha, \beta)}{\partial \, \beta}
\end{align*}

where $q$ can be derived from the definition of default correlation:

\begin{equation}
\rho = \frac{q - p p_{0}}{\sqrt{p (1 - p)} \sqrt{p_{0} (1 - p_{0})}}
\end{equation}

After a bit of algebra:

\begin{align*}
p_{0} = \frac{1}{1 + e^{-\alpha_{0}} (\frac{1 + e^{\alpha}}{1+e^{\alpha + \beta}})^N}\\
p = \frac{1}{1 + e^{-\alpha}} \left[1 + e^{\alpha_{0}} (e^{\beta} - 1) \frac{(1 + e^{\alpha + \beta})^{N-1}}{(1 + e^{\alpha})^{N} + e^{\alpha_{0}} (1 + e^{\alpha + \beta})^{N}} \right]\\
q = p_{0} \frac{1}{1 + e^{-\alpha - \beta}}
\end{align*}

We have three equations, relating our three empirical estimates, $p$, $p_{0}$ and $\rho$, with the three variables of the model, $\alpha$, $\alpha_{0}$ and $\beta$. After a bit of algebra, the relations above can be inverted explicitly:

\begin{align*}
\alpha_{0} = (N-1) \log\left(\frac{1-p_{0}}{p_{0}}\right) + N \log\left(\frac{p_{0} - q}{1 - p_{0} - p + q}\right)\\
\alpha = \log\left(\frac{p - q}{1 - p_{0} - p + q}\right)\\
\beta = \log\left(\frac{q}{p_{0} - q} \frac{1 - p_{0} - p + q}{p - q}\right)
\end{align*}

The probability distribution for the losses of the Dandelion model can then be calculated explicitly:

\begin{equation}
P\left(\ell = \sum\limits_{i=1}^N l_{i}\right) = \dfrac{1}{Z} \sum\limits_{l_{0},l_{1},\cdots,l_{N} | l_{1} + \cdots + l_{N} = \ell} \exp{ \left(\alpha_{0} l_{0} + \alpha \sum\limits_{i = 1}^N l_{i} + \beta \sum\limits_{i=1}^N l_{0} l_{i}\right)}
\end{equation}

Summing over $l_{0}$:

\begin{equation}
P\left(\ell = \sum\limits_{i=1}^N l_{i}\right) = \dfrac{1}{Z} \sum\limits_{l_{1},\cdots,l_{N} | l_{1} + \cdots + l_{N} = \ell} \exp{\left(\alpha \ell\right)} + \exp{\left(\alpha_{0} + \ell (\alpha + \beta)\right)}
\end{equation}

There are $\binom{N}{\ell}$ equal terms in the sum, finally giving:

\begin{equation}
P\left(\ell = \sum\limits_{i=1}^N l_{i}\right) = \dfrac{1}{Z} \binom{N}{\ell} \left(\exp{(\alpha \ell)} + \exp{(\alpha_{0} + \ell (\alpha + \beta))} \right)
\end{equation}

where $Z$, $\alpha$, $\alpha_{0}$ and $\beta$ depend on $p$, $p_{0}$ and $\rho$ as shown before.

\section{The Diamond model}

The Diamond model is defined by:

\begin{equation}
Z = \sum_{l_{1},l_{2},\ldots,l_{N}} \exp{ \left(\alpha \sum\limits_{i=1}^N { l_{i}} + \beta \sum\limits_{i > j}{ l_{i} l_{j}}\right)}
\end{equation}

Since

\begin{equation}
\ell^2 = \sum\limits_{i=1}^N { l_{i}} \sum\limits_{j=1}^N { l_{j}} = \sum\limits_{i=1}^N { l_{i}^2} + 2 \sum\limits_{i > j}{ l_{i} l_{j}}
\end{equation}

We can rewrite the partition function as:

\begin{equation}
Z = \sum_{l_{1},l_{2},\ldots,l_{N}} \exp{ \left(\alpha \ell + \beta \frac{\ell(\ell - 1)}{2}\right)}
\end{equation}

Due to homogeneity, we can sum over states with the same $\ell$, with a degeneracy of $\binom{N}{\ell}$:

\begin{equation}
Z = \sum_{\ell=0}^N \binom{N}{\ell} \exp{ \left(\left(\alpha - \frac{\beta}{2}\right) \ell + \frac{\beta}{2} \ell^2\right)}
\end{equation}

And the corresponding probability distribution for the losses will be:

\begin{equation}
P(\ell = \sum\limits_{i=1}^N l_{i}) = \binom{N}{\ell} \frac{ \exp{ \left((\alpha - \frac{\beta}{2}) \ell + \frac{\beta}{2} \ell^2\right)}}{Z}
\end{equation}

We can relate the empirical data, $p$ and $\rho$ to the model parameters $\alpha$ and $\beta$ by:

\begin{align*}
p = \frac{\sum\limits_{i = 1}^N \langle l_{i} \rangle}{N} = \frac{1}{N} \frac{\partial \, \log Z(\alpha, \beta)}{\partial \, \alpha}\\
q = \frac{\sum\limits_{i > j} \langle l_{i} l_{j} \rangle}{\frac{N (N-1)}{2}} = \frac{2}{N (N-1)} \frac{\partial \, \log Z(\alpha, \beta)}{\partial \, \beta}
\end{align*}

Then:

\begin{align*}
p = \frac{1}{Z N} \sum_{\ell=0}^N \binom{N}{\ell} \ell \exp{ \left((\alpha - \frac{\beta}{2}) \ell + \frac{\beta}{2} \ell^2\right)}\\ q =\frac{2}{Z N (N-1)} \sum_{\ell=0}^N \binom{N}{\ell} \frac{1}{2} \ell (\ell-1) \exp{ \left((\alpha - \frac{\beta}{2}) \ell + \frac{\beta}{2} \ell^2\right)}
\end{align*}

\end{document}